\documentclass{IEEEtran}
\usepackage{cite}
\usepackage{amsmath,amssymb,amsfonts}
\usepackage{algorithmicx}
\usepackage{graphicx}
\usepackage{float}
\usepackage{textcomp}
\usepackage{subfigure}
\usepackage{comment}
\usepackage{color,xcolor}
\usepackage{lineno}
\usepackage{url} 
\usepackage{booktabs}
\def\BibTeX{{\rm B\kern-.05em{\sc i\kern-.025em b}\kern-.08em
    T\kern-.1667em\lower.7ex\hbox{E}\kern-.125emX}}
\begin{document}
\title{Channel Parameter Estimation and Characteristic Analysis at THz Bands in Hybrid Near-Field/Far-Field Scenarios}
\author{Hengtai Chang, \IEEEmembership{Member, IEEE}, Cheng-Xiang Wang, \IEEEmembership{Fellow, IEEE}, Cunhua Pan, \IEEEmembership{Senior Member, IEEE}, Jian Sun, \IEEEmembership{Member, IEEE}, Bingchang Hua, Yongchao He, and el-Hadi M. Aggoune,~\IEEEmembership{Life Senior Member,~IEEE}
\thanks{This work was supported by the National Key R\&D Program of China under Grant 2023YFB2905100, the Research Fund of National Mobile Communications Research Laboratory, Southeast University, under Grant 2026A05, the Shandong Provincial Natural Science Foundation under Grant ZR2025MS1096, and the AI and Sensing Technologies Center, University of Tabuk, KSA. (\emph{Corresponding Author: Cheng-Xiang Wang}).}
\thanks{H. Chang and J. Sun are with School of Information Science and Engineering, Shandong University, Qingdao 266237, China (e-mail: \{changht, sunjian\}@sdu.edu.cn).}
\thanks{C.-X. Wang is with the National Mobile Communications Research Laboratory, School of Information Science and Engineering, Southeast University, Nanjing 211189, China and also with the Purple Mountain Laboratories, Nanjing 211111, China (email: chxwang@seu.edu.cn).}
\thanks{C. Pan and Y. He are with the National Mobile Communications Research Laboratory, School of Information Science and Engineering, Southeast University, Nanjing 211189 (email: \{cpan, heyongchao\}@seu.edu.cn).}
\thanks{B. Hua is with Purple Mountain Laboratories, Nanjing 211111, China (e-mail: huabingchang@pmlabs.com.cn).}
\thanks{E. M. Aggoune is with AI and Sensing Technologies Research Center, University of Tabuk, Tabuk 47315/4031, Saudi Arabia (e-mail: haggoune@ut.edu.sa).}
}
\markboth{IEEE Transactions on Vehicular Technology,~Vol.~xx, No.~xx, MONTH 2026}%
{Shell \MakeLowercase{\textit{et al.}}: Bare Demo of IEEEtran.cls for IEEE Journals}
\maketitle
\begin{abstract}
The terahertz (THz) band is a promising candidate for high-data-rate transmission in future wireless communication systems, where large-scale antenna arrays are deployed to combat the severe propagation loss. In such configurations, near-field and far-field multipath components (MPCs) coexist in the channel, which challenges conventional parameter estimation methods. This paper presents a 260--380 GHz virtual uniform linear array (ULA) measurement campaign in an indoor scenario, and proposes a hybrid-field space-alternating generalized expectation-maximization (SAGE) algorithm that discriminatively estimates near-field and far-field MPCs via a visible region (VR)-aided weighted wideband Bayesian phase-curvature classification, while tracking spatial birth-death phenomena through VR estimation. Simulations with known ground truth validate the effectiveness of the proposed classifier and the hybrid wavefront modeling.  Measurement-based characterization reveals that near-field MPCs account for over 90\% of total MPCs at a 2 m distance, the scatterer distances and invisible region lengths follow lognormal and exponential distributions, respectively, and spatial non-stationarity intensifies with increasing frequency and decreasing distance. These findings provide quantitative guidelines for THz channel modeling and system design.

\end{abstract}


\section{Introduction}
\label{sec:introduction}
\IEEEPARstart{T}{he} sixth-generation (6G) wireless systems are promising to achieve the vision of ``all-spectra, global-coverage, full-application, and strong security \cite{Wang2023}.'' Among the key technologies envisioned to realize this vision, terahertz (THz) communications stand out due to their potential to offer ultra-high data rates, leveraging the vast available bandwidth in the THz frequency band \cite{Akyildiz2014}. The THz band, typically spanning from 0.1 to 10 THz, presents unique propagation characteristics that differentiate it from lower-frequency bands such as microwave and millimeter-wave (mmWave) bands \cite{Molisch2021,Wang2024}. These characteristics include high isotropic free-space path loss \cite{Rappaport2019}, strong molecular absorption \cite{Jornet2011}, severe frequency selectivity \cite{Han2015} and non-stationarity  \cite{Wang2022}, etc. To enhance the effectiveness and reliability of THz communication systems, the introduction of large-scale antenna arrays is crucial \cite{Han2022}. On one hand, THz waves suffer from severe path loss due to their high frequency and short wavelength \cite{Petrov2020}. Large-scale antenna arrays can provide high beamforming gains, effectively compensating for high path loss and ensuring reliable signal transmission and reception \cite{Han2017}. On the other hand, large-scale antenna arrays enable spatial multiplexing, allowing multiple data streams to be transmitted simultaneously over the same frequency band \cite{Bian2021}. Large-scale antenna arrays significantly increase the spectral efficiency of the system, which is important for achieving high data rates envisioned for 6G. It is well known that precise analysis and characterization of wireless channels are the foundations of follow-up technical research. As a result, the study of THz wireless propagation channels  is fundamental for the development of reliable THz communication systems and their applications \cite{Han2022}.

Recent research has seen significant efforts in THz channel measurement and characterization. Current THz channel measurements employ three primary methodologies to address distinct propagation challenges. {Frequency-domain vector network analyzer (VNA)-based systems} serve as the benchmark for high-precision static channel characterization, leveraging S-parameter measurements to resolve path loss and multipath components (MPCs). For instance, \cite{Abbasi2020} utilized  radio-over-fiber-extended VNA at 140-300 GHz to quantify double-directional urban path loss over 100 m distances, while \cite{Chen2021a,Chen2021b} mapped 140 GHz angular spreads in indoor environments using three-dimensional (3D) horn-antenna scans. {Time-domain sliding correlators (TDS)} enable dynamic channel analysis through PN-sequence correlation, supporting mobility and real-time applications. For example, \cite{Xing2021,Ju2021} captured spatial consistency in 142 GHz urban microcells using phased-array sounders, and \cite{Eckhardt2019} characterized 300 GHz data center blockage dynamics with sub-ns temporal resolution. {THz-TDS systems} focus on ultra-wideband material interactions and near-field effects, where \cite{Priebe2013} extracted 300 GHz rough-surface scattering coefficients via pulsed measurements. These approaches collectively reveal THz-specific propagation traits, including {distance-dependent path loss exponents} (line-of-sight (LoS): $\sim$2.0 at 300 GHz \cite{Chen2021b}; non-LoS (NLoS): 2.5-4.0 \cite{Abbasi2021}), {multipath sparsity} (root-mean-square delay spreads $<$30~ns indoors \cite{Chen2021a} vs. $>$100 ns outdoors \cite{Xing2021b}), and \textit{material-dependent reflectivity} (0.1-0.8 for concrete at 300 GHz \cite{Priebe2013}). Despite progress, critical gaps persist in standardized $>$300~GHz multi-antenna measurements and characterization \cite{NextG2021}. The European Telecommunications Standards Institute (ETSI) Group Report (GR) THz 002 report provides an overview of the current regulatory landscape for THz frequency bands. It identifies the 260-380 GHz frequency band as a key area for research and standardization \cite{ETSI}. In addition, the International Telecommunication Union (ITU) Radio regulations have identified certain frequency bands within 275-450 GHz for use by administrations for land mobile and fixed service applications~\cite{ITU-R}. 

Despite advancements in THz channel measurements, the development of efficient and accurate channel parameter estimation algorithms for THz channels remains an open challenge. Traditional algorithms, such as Bartlett spectrum \cite{Bartlett1948}, Capon spectrum \cite{Capon2005}, unitary estimation of signal parameter via rotational invariance techniques (ESPRIT) \cite{Zhang2017}, multiple signal classification (MUSIC) \cite{Guo2017}, Richter's maximum likelihood (ML) estimation (RiMAX)\cite{Richer2003}, and expectation maximization (EM)\cite{K1996} designed primarily for far-field assumption and lower-frequency bands, often fail to adequately capture the unique features of large-scale antenna array THz channels. In addition, building upon the EM algorithm, the space-alternating generalized expectation-maximization (SAGE) algorithm \cite{Fleury1999} effectively reduces complexity by partitioning channel parameters into several subsets.
As the number of antennas increases, forming a massive multiple-input multiple-output (MIMO) system, the spherical wavefront caused by near-field effect can be observed \cite{Zhang2019}. Under these conditions, a plane wavefront (far-field) model no longer matches the measured wavefront, causing significant estimation bias or estimation failure \cite{Yi2018}. In the works of Ma \cite{Ma2020} and Zhang \cite{Zhang2018}, the MUSIC algorithm was enhanced by incorporating near-field scatterers. Yet, this approach still falls short in addressing large bandwidth and spatial non-stationary characteristics. Meanwhile, \cite{Yin2017} introduced an algorithm grounded in the EM principle to estimate wideband 3D near-field channel parameters. In \cite{HanSAGE}, a direction-scan sounding (DSS)-o-SAGE algorithm that effectively addresses phase instability and reduces computational complexity in channel parameter estimation for mmWave and THz bands was developed for direction-scan sounding. In addition, Zhou et al. \cite{ZhouSAGE} introduced a new SAGE algorithm which focuses on estimating parameters for wideband spatial nonstationary wireless channels that incorporate antenna polarization. This highlights the need for innovative algorithms that can effectively model and estimate large-scale antenna array channel parameters in the THz bands. However, the existing channel parameter estimation algorithms all focus on near-field or far-field environments. The hybrid-field environments are rarely studied. ``hybrid-field'' denotes a propagation scenario in which the wireless channel simultaneously contains both near-field and far-field MPCs, so that spherical-wave and planar-wave MPCs coexist. Recently, a deep-unfolded Bayesian learning approach was proposed in \cite{Hybrid2026} for wideband hybrid-field THz UM-MIMO channel estimation. However, it aims at recovering channel coefficients rather than extracting physically interpretable multipath parameters, and is validated only through simulations.

\indent To fill the above-mentioned research gaps, in this manuscript we conduct channel measurement with the virtual large scale antenna array at THz band (260--380 GHz) and develop a novel hybrid-field SAGE algorithm for parameter estimations of spatial non-stationary wireless channels at THz bands. Based on the channel measurement and hybrid-field parameter estimation, this paper also analyzes the influence of the communication environment and system configuration on near-field effects comprehensively. The main contributions and novelties of this paper are as~follows.
\begin{itemize}
	\item[$\bullet$] We propose an improved hybrid-field SAGE algorithm for channel parameter estimation under hybrid near-field/far-field conditions. A VR-aided weighted wideband Bayesian classifier is developed to discriminate between near-field and far-field MPCs, in which the visible region (VR) of each MPC is pre-detected and only the reliable elements within it contribute to the phase-curvature test, making the classification robust to spatial non-stationarity. The VRs of MPCs are further refined to capture their spatial birth-death behavior along the array.

	\item[$\bullet$] We validate the proposed algorithm through both simulations with known ground truth and 260--380 GHz channel measurements on a virtual large-scale ULA platform in an indoor scenario. Simulation results confirm the reliability of the proposed classifier and the advantage of hybrid wavefront modeling, while the measurement data demonstrate the algorithm's effectiveness in resolving hybrid-field MPCs in practical environments.

	\item[$\bullet$] We conduct an in-depth characterization of the large-scale array THz channel based on the proposed algorithm and measurements. The results reveal that near-field MPCs account for over 90\% of total MPCs at 2 m distance (380 GHz), that the scatterer distances and invisible region lengths follow lognormal and exponential distributions, respectively, and that spatial non-stationarity intensifies as the carrier frequency increases and the transmission distance decreases. These findings provide quantitative foundations for THz channel modeling and useful guidance for the design of large-scale-array THz communication systems in future 6G networks.
\end{itemize}

The rest of this paper is organized as follows. Section~\ref{THz Channel Measurements} introduces the virtual multi-antenna THz channel measurement setup, environment, and system calibration process for THz channel sounding.
In Section~\ref{Channel Parameter Estimation}, the hybrid-field SAGE algorithm and its parameter estimation workflow are presented.
Then, Section~\ref{Channel Characterization and Evaluation Using Real Measurement Data} analyzes THz channel properties including delay power spectral density (PSD), path loss, hybrid-field MPC characteristics, spatial 
-frequency correlations, and scatterer distributions.
Finally, conclusions are drawn in Section~\ref{Conclusions}.

\section{Virtual Multi-antenna THz Channel Measurements}
\label{THz Channel Measurements}
\subsection{Channel Sounding Systems}
\indent The Ceyear VNA-based frequency-domain channel sounder is used to conduct THz channel measurements in this work. The THz channel sounding system is composed of three parts, as shown in Fig. 1, including the computer (PC) as the control platform, the displacement platform to form the virtual antenna array, and the sounding platform to get the channel response. The sounding platform consists of THz transmitter (Tx) module, Rx module, and the Ceyear 3672C VNA. The VNA outputs local oscillator (LO) and radio frequency (RF) references. The RF signal is up-converted $\times 27$ to reach the THz carrier frequency. The LO signal is up-converted $\times 24$. Reference and test IFs from Tx and Rx are returned to the VNA, whose ratio gives the THz channel transfer function (CTF) \cite{AWPL2025}. The calculated channel response contains not only the wireless channel but also the device, cables, and waveguides.
\begin{figure}
    \centering
    \includegraphics[width=1\linewidth]{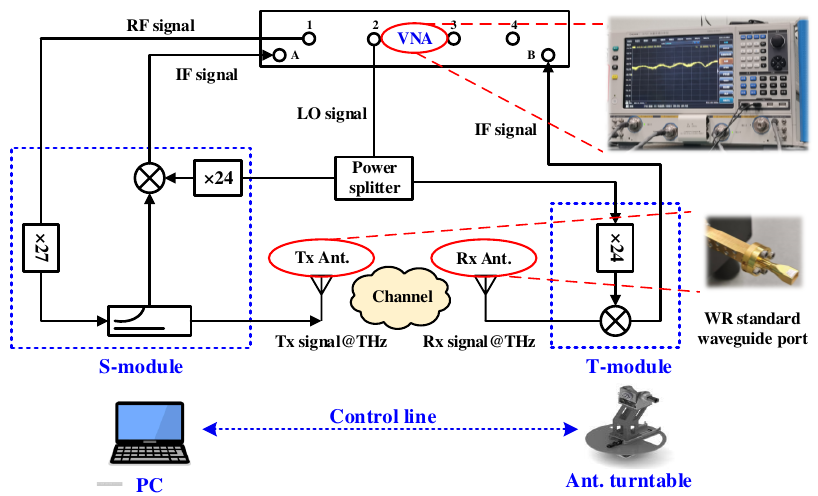}
    \caption{THz channel measurement system setup.}
    \label{fig:enter-label}
\end{figure}
A VNA-based channel sounder operating at 260, 330, and 380 GHz is used for the indoor THz measurements. Each frequency band covers a 20 GHz sounding bandwidth sampled at 1 001 points with 20 MHz spacing, $\Delta f_m = 20\ \text{MHz}$, yielding 50 ps time resolution, $\tau_m=1/\Delta f=50\ \text{ns}$, and 1.5 cm path resolution, $\Delta L=1.5\ \text{cm}$. Accordingly, the maximum detectable path length is  $L_m=15\ \text{m}$, which is sufficient to meet the needs of conference room channel measurements described below. Both the Tx and Rx are equipped with horn antennas, with the antenna gain of 25 dBi, and a half-power beamwidth (HPBW) of 15 degrees, allowing for a high measurement dynamic range. The Rx is mounted on a stepping table, while the Tx needs to be positioned at different points within the conference room. Additionally, the test signal power is 0.5 mW, and the background noise of the measurement platform is -145 dBm. The detailed parameters of the indoor THz measurement system are summarized in Table \ref{Tab1}.

\begin{table}[!ht]
\centering
\caption{Measurement System Setup}
\begin{tabular}{lcc}
\toprule
\textbf{Parameter} & \textbf{Symbol} & \textbf{Value} \\
\midrule
RF signal frequency & $f_{\mathrm{RF}}$ & 9.63--14.81 GHz \\
LO signal frequency & $f_{\mathrm{LO}}$ & 10.84--16.68 GHz \\
IF signal frequency & $f_{\mathrm{IF}}$ & 279 MHz \\
Start Sounding Frequency & $f_{\mathrm{start}}$ & 260/330/380 GHz \\
Stop Sounding Frequency & $f_{\mathrm{end}}$  & 280/350/400 GHz \\
Bandwidth & $B_{\mathrm{u}}$ & 20 GHz \\
Virtual Antenna Number & $M$ & 128 \\
Sweep Frequency Interval & $\Delta f_m$ & 20 MHz \\
Average Noise Floor & $P_{\mathrm{n}}$ & $-$145 dBm \\
Transmission Power & $P_{\mathrm{in}}$ & 0.5 mW \\
HPBW of Tx Antenna& $\mathrm{HPBW}_{\mathrm{T}}$ & 15$^{\circ}$ \\
HPBW of Rx Antenna& $\mathrm{HPBW}_{\mathrm{R}}$ & 15$^{\circ}$ \\
Antenna Gain at Tx & $G_{\mathrm{T}}$ & 25 dBi \\
Antenna Gain at Rx& $G_{\mathrm{R}}$ & 25 dBi \\
Time Delay Resolution & $\Delta\tau$ & 50 ps \\
Path Length Resolution & $\Delta L$ & 1.5 cm \\
Maximum Excess Delay & $\tau_{\mathrm{m}}$ & 50 ns \\
Maximum Path Distance & $L_{\mathrm{m}}$ & 15 m \\
\bottomrule
\label{Tab1}
\end{tabular}
\end{table}

\subsection{Channel Measurement Environment and Setup}
Channel measurements were conducted in a typical conference room at the Purple Mountain Laboratories, with the layout of the room and diagram shown in Fig. 2. The conference room is a rectangular area of 10 m × 5.6 m, with a small conference table with the size of 1.4 m × 0.6 m and placed in the center of the room, at a height of 0.74 m. In the 10 m × 5.6 m conference room, six seats ring the central table. A TV is mounted on the right-hand wall where different panels meet. Behind the Tx is a glass wall, while in front of it is a wooden wall. The wall directly opposite the conference room door serves as the boundary for the location of the Rx, which is divided into a plastic wall and a wooden wall, while the other two walls are glass walls. The conference room has a narrow and elongated environment. Smooth wall finishes and the elongated layout generate rich reflective paths. As mentioned before, the maximum detectable path length of the measurement system is 15 meters, which is larger than the size of the conference room. Therefore, rich scattering and reflection paths can be recorded during the measurement.

To simulate the indoor communication scenario, an Rx with a half-power beamwidth of 15 degrees is fixed near the midpoint at one end of the conference room, while a Tx with the same half-power beamwidth is placed at the other end of the room. Both the Tx and Rx are set at a height of 0.9 m, which is above the height of the tables and chairs in the room. In performing channel measurements, the Rx is positioned at location "Rx" in Fig. 2, and the Tx is placed at locations Tx1-25 in Fig. 2, with a vertical distance of 0.5 m and a horizontal distance of 0.3 m between the Tx positions. To facilitate measurement, the main beams of the Tx and Rx are always aligned and parallel to the direction of the conference room, ensuring that all Tx positions are within Rx beam range and that there are no obstacles blocking the LoS path. During the channel measurements, the Rx is placed on a rotating platform while the Tx remains stationary, and the rotating platform is moved by half a wavelength each time, resulting in a total of 128 points to achieve a virtual large-scale uniform linear array (ULA) with 128 antenna elements. The corresponding antenna aperture is 6.4 cm and the near-field range is about 6.1 m according to the calculation of Rayleigh distance. Therefore, the whole measurement environment contains both near-field and far-field regions. Besides, we limit the rotation speed so the Tx–Rx geometry stays effectively static while the VNA sweeps, guaranteeing precise amplitude and phase acquisition. In order to ensure that the virtual antenna array is equivalent to the real antenna array in the measurement, the environment remains unchanged during the measurement process.

\begin{figure}
    \centering
    \includegraphics[width=1\linewidth]{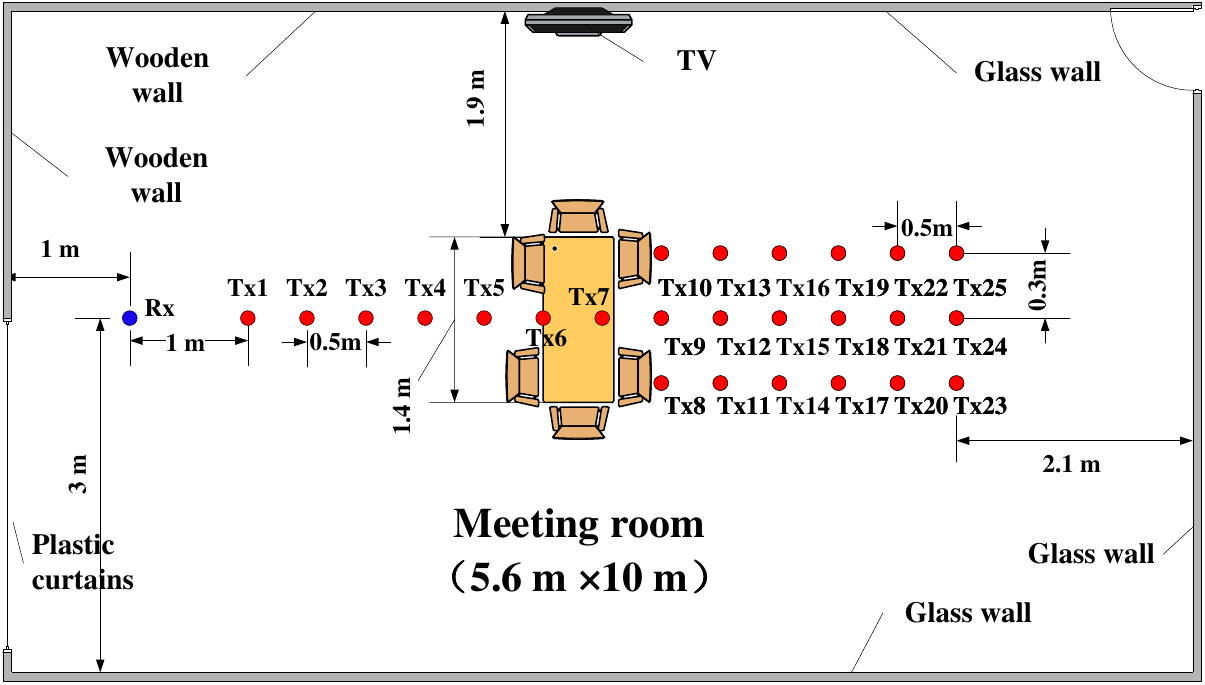}
    \caption{Layout of indoor channel measurement scenario.}
    \label{fig:enter-label}
\end{figure}
\subsection{Channel Sounding System Calibration}
Prior to measurements, we calibrate the setup to remove system responses. All data are then collected via VNA-based frequency-domain sweeping. The VNA can directly output the values of signal amplitude and phase (i.e., S parameters) and can directly display the amplitude values of multipath through its time-domain function. The dynamic range can reach up to 140 dB in the frequency range of 260--380 GHz \cite{AWPL2025}. To obtain accurate S21 parameters, it is essential to perform a back-to-back calibration for the THz sounding system to eliminate the impact of the frequency domain response caused by antennas, VNA, and cables on the measurement data. By directly connecting the T receiving module and the S transmitting module through a 50 dB attenuator, we can obtain the overall frequency domain response of the measurement system itself. The impact can then be eliminated through amplitude and phase normalization processing in the VNA. The direct connection calibration operation is simple. The calibrated CTF for each antenna pair can be expressed~as
\begin{align}
H(f) = \frac{{{S_{21}}(f)}}{{{H_{{\rm{Tx}}}}(f){H_{{\rm{sys}}}}(f){H_{{\rm{Rx}}}}(f)}}
\end{align}
where ${H_{{\rm{Tx}}}}(f)$  and  ${H_{{\rm{Rx}}}}(f)$ represent the frequency domain responses of the antennas on the receiving and transmitting sides, respectively, and   ${H_{{\rm{sys}}}}(f)$ represents the frequency domain response caused by the VNA and cables. The VNA sequentially measures the channel responses at each frequency point across the bandwidth, while the CTF for the entire measurement bandwidth is the collection of all frequency~points.

\section{Channel Parameter Estimation}
\label{Channel Parameter Estimation}
\subsection{Signal Model}
Let us consider a wideband single-input multiple-output (SIMO) channel in the hybrid-field environment. The CTF of the multipath channel can be represented by a matrix ${\bf{H}} \in {\mathbb{C}^{M \times K}}$ and each element in ${\bf{H}}$ can be written as the superposition of near-field MPCs and far-field MPCs
\begin{align}
{{{H}}{[m,k]}} &= \sum\limits_{l \in {\cal C}_N} {{\alpha _l} \cdot {\xi _{l,m}} \cdot {e^{-j\frac{{2\pi }}{{{\lambda _k}}}\left( {{d_{{\rm{Rx,}}l,m}} - {d_{{\rm{Rx,}}l}}} \right)}} \cdot {e^{ - j2\pi {\tau _l}{f_k}}}} \nonumber\\&+\sum\limits_{l \in {\cal C}_F} {{\alpha _l} \cdot {e^{-j\frac{{2\pi }}{{{\lambda _k}}}{\delta _d}(m - 1)\sin {\theta _l}}} \cdot {e^{ - j2\pi {\tau _l}{f_k}}}} 
\end{align}
where $m$ is the antenna element index, $k$ is the frequency point index, $\delta_d$ is the antenna spacing, $l$ is the MPC index, $\alpha _l$  is the complex gain of the $l$-th MPC on the first antenna element,  $\xi _{l,m}$ is the birth-death factor of the $l$-th MPC on the $m$-th antenna element, in which 1 indicates visible, and 0 indicates invisible, $\tau _l$  is the time delay of the $l$-th MPC,  ${f_k}$ is the frequency of $k$-th carrier in the bandwidth, and  ${\lambda _k} = c/{f_k}$ is the wavelength of ${f_k}$. The distance difference considering spherical wavefront is denoted by ${d_{{\rm{Rx,}}l,m}} - {d_{{\rm{Rx,}}l}}$ , where ${d_{{\rm{Rx,}}l,m}}$ is the propagation distance from last-bounce scatterer of the $l$-th MPC to the $m$-th Rx antenna element and ${d_{{\rm{Rx,}}l}}$ is the propagation distance from last-bounce scatterer of the $l$-th MPC to the first Rx antenna element. Here we set the broadside of Rx antenna array is the azimuth direction of 0 degree and the azimuth angle of arrival is ${\theta _l}$. Thus, the distance difference can be derived as ${d_{{\rm{Rx,}}l,m}} - d_{{{\rm{Rx,}}l}} \approx {\delta _d}(m - 1)\sin {\theta _l} + \frac{{{\delta _d}^2{{(m - 1)}^2}}}{{2{d_l}}}$  according to Taylor expansion. The parameters of near-field MPC that need to be extracted from measurement data can be represented by the set ${{\bf{\Gamma }}_{N,l}} = \left\{ {\Re \left\{ {{\alpha _l}} \right\},} \right.{\rm{ }}\Im \left\{ {{\alpha _l}} \right\},{\rm{ }}\left. {{d_{{\rm{Rx}},l}},{\tau _l},{\theta _{l}},{\xi _{l,s}},{\xi _{l,e}}} \right\} \in {\mathbb{R}^7}$, in which ${\xi _{l,s}}$  and ${\xi _{l,e}}$  are the start antenna index and end antenna index of the VR on the Rx antenna array, respectively. As for the far-field MPC, the distance parameter $d_{{\rm{Rx}},l}$ and birth-death parameters ${\xi _{l,s}}$  and ${\xi _{l,e}}$ are no longer needed. The extracted parameter set can be simplified to ${{\bf{\Gamma }}_{F,l}} = \left\{ {\Re \left\{ {{\alpha _l}} \right\},} \right.{\rm{ }}\Im \left\{ {{\alpha _l}} \right\},{\rm{ }}\left. {{\tau _l},\ {\theta _l}} \right\} \in {\mathbb{R}^4}$. Therefore, the CTF $\mathbf{H}$ can be rewritten as
\begin{align}
{\bf{H}}\left( {{\boldsymbol{f}},{\bf{\Gamma }}} \right) = \sum\limits_{{l \in {\cal C}_N}} {{\bf{H}}\left( {{\boldsymbol{f}},{{\bf{\Gamma }}_{N,l}}} \right)} + \sum\limits_{{l \in {\cal C}_F}} {{\bf{H}}\left( {{\boldsymbol{f}},{{\bf{\Gamma }}_{F,l}}} \right)} 
\end{align}
where ${\boldsymbol{f}} = \left[ {{f_1},{f_2},...,{f_k}} \right]$  represents the $K$ frequency points and ${\bf{\Gamma }} = \left\{ {{\bf{\Gamma }}_{N,{\cal C}_F}},{{\bf{\Gamma }}_{F,{\cal C}_N}} \right\}$  contains parameters of all MPCs. Considering the inevitable measurement noise, the measured CTF can be expressed as
\begin{align}
{\bf{Y}}\left( {\boldsymbol{f}} \right) = {\bf{H}}\left( {{\boldsymbol{f}},{\bf{\Gamma }}} \right) + \sqrt {\frac{{{\sigma _n}^2}}{2}} {\bf{n}}\left( {\boldsymbol{f}} \right)
\end{align}
where ${\bf{n}}\left( {\boldsymbol{f}} \right)$  denotes the noise matrix, in which each element is assumed to follow an independent and identically distributed standard complex Gaussian distribution with zero mean and unit variance. Besides, ${\sigma _n}^2$ represents the variance of the noise~matrix.
\subsection{Parameter Estimation Algorithm}
Here, we leverage the maximum likelihood criterion to extract parameters of each MPC. The expectation of the log-likelihood function can be obtained by 
\begin{align}
\label{Likely}
{\mathbb{E}}[{\cal L}({\bf{\Gamma }};{\bf{Y}}({\boldsymbol{f}}))] =  &- {\rm{ }}2\ln \left( {\pi \sigma _n^2} \right) - \frac{1}{{\sigma _n^2}}{\rm{ }} \cdot \big \|{\mathop{\rm vec}\nolimits} \{ {\bf{Y}}({\boldsymbol{f}})\}  \nonumber\\ &- {\mathop{\rm vec}\nolimits} \{ {\bf{H}}({\boldsymbol{f}};{\bf{\Gamma }})\}\big \| {^2}
\end{align}
in which ${\mathbb{E}}[ \cdot ]$  refers to the expectation operation, and ${\rm vec}\{\cdot\}$ represents the vectorization operation. The parameter set is estimated via maximum-likelihood maximization in (5), and can be obtained by
\begin{align}
\label{ML2}
\widehat {\bf{\Gamma }} = \mathop {{\mathop{\rm argmax}\nolimits} }\limits_{\bf{\Gamma }} \{ {\mathbb{E}}[{\cal L}({\bf{\Gamma }};{\bf{Y}}({\boldsymbol{f}}))]\} .
\end{align}

\indent Due to the high dimensionality of $\boldsymbol{\Gamma}$, direct maximization of (\ref{ML2}) becomes computationally prohibitive. Instead, we adopt a successive interference cancellation (SIC) framework introduced in \cite{Fessler1994,Xue2003}, which proceeds under the orthogonality assumption among individual MPCs. Within the SIC framework, components are extracted sequentially according to descending power, thereby mitigating the leakage from dominant MPCs to weaker ones. Owing to the high delay and angular resolution inherent to wideband massive MIMO channels, any pair of MPCs whose delay or angular separation exceeds the system’s resolution limit can be treated as mutually orthogonal \cite{Fleury1999}.
Consequently, this work employs the SIC technique to retrieve each MPC’s parameters in a sequential manner. Specifically, the $l$-th component is obtained by maximizing 
\begin{equation}
	\label{ML3}	
	\hat{\boldsymbol{\Gamma}}_l   = \mathop{\mathrm{argmax}}\limits_{\boldsymbol{\Gamma}_l }\big{\{} \textbf{E}[{\cal L} (\boldsymbol{\Gamma }_l;\textbf{Y}_l (\boldsymbol{f}) )]\big{ \}}
\end{equation}
where $\textbf{Y}_l (\boldsymbol{f}) \in \mathbb{C}^{M\times K} $ represents the expectation of the $l$-th MPC from the measured CTF and can be calculated by \cite{Yi2018}
\begin{equation}
	\textbf{Y}_l (\boldsymbol{f} )=\begin{cases} \textbf{Y} (\boldsymbol{f} )
		& \text{ if } l=1.  \\
		\textbf{Y} (\boldsymbol{f} )-\sum_{l^{'}=1}^{l-1} \textbf{H} (\boldsymbol{f};\hat{\boldsymbol{\Gamma  }}_{l^{'}} )
		& \text{ if } l\in \left [ 2,\cdots,L \right ] .
	\end{cases}
\label{CTF_NS}
\end{equation}\par
The strongest MPC is first extracted from $\textbf{Y}(\boldsymbol{f})$. The reconstructed channel response given by (\ref{CTF_NS}) is then subtracted from the $\textbf{Y}(\boldsymbol{f})$ and the residual part is used to estimate the subsequent MPC. The process continues until all $L$ significant MPCs have been identified. Furthermore, it can be proved that maximizing the log-likelihood function in (\ref{ML3}) is equivalent to maximizing the objective function $z(\boldsymbol{f} ;\boldsymbol{\Gamma} _l^{\alpha} )$ as \cite{Xue2016}
\begin{equation}
	\label{ML4}
	\hat{\boldsymbol{\Gamma}}_l^{\alpha}   = \mathop{\mathrm{argmax}}\limits_{\boldsymbol{\Gamma}_l^{\alpha} }\big{\{}z(\boldsymbol{f} ;\boldsymbol{\Gamma} _l^{\alpha} )\big{\}}
\end{equation}
\begin{equation}
{\boldsymbol{\Gamma}} _l^{\alpha}= 
\begin{cases}
{\boldsymbol{\Gamma}} _{{\cal C}_{N},l}^{\alpha}&\text{ if } l\in {\cal C}_{N}\\
{\boldsymbol{\Gamma}} _{{\cal C}_{F},l}^{\alpha}&\text{ if } l\in {\cal C}_{F}  
\end{cases}
\end{equation}
where ${\boldsymbol{\Gamma}} _l^{\alpha}$ represents the parameter set of the $l$-th MPC without ${\boldsymbol{\alpha}}_l$, in which ${\boldsymbol{\Gamma}} _{{\cal C}_{N},l}^{\alpha}= \left\{ \right.\left. {{d_{{\rm{Rx}},l}},{\tau _l},{\theta _l},{\xi _{l,s}},{\xi _{l,e}}} \right\} \in {\mathbb{R}^5}$, and ${\boldsymbol{\Gamma}} _{{\cal C}_{F},l}^{\alpha}= \left\{ \right.\left. {{\tau _l},{\theta _l}} \right\} \in {\mathbb{R}^2}$, and $z(\boldsymbol{f} ;\boldsymbol{\Gamma} _l^{\alpha})$ can be calculated~by
\begin{small}
\begin{equation}
	\label{z}
		z(\boldsymbol{f} ;\boldsymbol{\Gamma} _l^{\alpha})=\left \| \boldsymbol{C}^{\rm H}_l \textbf{Y}_l(\boldsymbol{f} )\boldsymbol{D}(\tau_l) \right \| 
\end{equation}
\end{small}%
\noindent where $[\cdot]^{H}$ denotes the conjugate transpose operation, and $\boldsymbol{C}_l$ denotes the steering vector of the Rx antenna array
\begin{align}
    \boldsymbol{C}_l=
    \begin{cases}
        \begin{array}{ll}\left[  \right.1, e^{\frac{j2\pi}{\lambda}({\delta _d}\sin {\theta _l} + \frac{{{\delta _d}^2}}{{2{d_l}}})}, ..., \\\quad \ e^{\frac{j2\pi}{\lambda}({\delta _d}(M - 1)\sin {\theta _l} + \frac{{{\delta _d}^2{{(M - 1)}^2}}}{{2{d_l}}})}  \left. \right] \odot {\boldsymbol{\xi }}_{l}
        \end{array}&\text{ if } l\in {\cal C}_{N}\\
        \left[  1, e^{\frac{j2\pi}{\lambda}({\delta _d}\sin {\theta _l})}, ..., e^{\frac{j2\pi}{\lambda}({\delta _d}(M - 1)\sin {\theta _l} )}  \right]&\text{ if } l\in {\cal C}_{F}
    \end{cases}
\end{align}
where ${\boldsymbol{\xi }}_{l}$ is a mask vector defining the VR of $l$-th MPC, i.e., 
\begin{align}
\label{Mask}
{\boldsymbol{\xi }}_{l}=\left[ {\bf{O}}_{{\xi _{l,s}}-1},{\bf E}_{{\xi _{l,e}}-{\xi _{l,s}}+1},{\bf{O}}_{{M}-{\xi _{l,e}}}\right ]
\end{align}
where ${\bf{O}}_n$ and ${\bf{E}}_n$ denote the all-zero vector and all-one vector with $n$ entries. Besides,  $\boldsymbol{D}(\tau_l)$ is the delay matrix defined by
\begin{align}
	\boldsymbol{D}(\tau_l)=\left[ e^{j2\pi f_1 \tau_l} ,e^{j2\pi f_2 \tau_l},...,e^{j2\pi f_K \tau_l}\right].
\end{align}
\subsection{Hybrid-field Parameter Estimation Process}
To ease the computational load, a coarse-to-fine search strategy is leveraged in the initialization process. The coarse stage first computes delays and angles under a narrowband and far-field model. Consequently, the parameter space of delay and angle meets the orthogonal stochastic measure (OSM) condition \cite{Krim1996}, enabling original SAGE processing.  The coarse-to-fine search process ignores the birth and death of MPCs in the spatial domain, indicating that $\boldsymbol{\xi}_{l}$ is set to an all-one matrix.  In the next step, delay and angle parameters are estimated based on the ML principle, while the spatial birth-death behavior of each MPC is pre-detected in the classification stage and refined in the subsequent birth-death identification step.
\subsubsection{Parameter coarse search with narrowband and far-field assumption}
The initial delay and angle of arrival (AoA) of the $l$-th MPC will be estimated at this step. The initial estimated  time delay of the $l$-th MPC, i.e., $\hat{\tau}_l^{\text{init}}$ can be calculated by 
\begin{equation}
	\label{tau-init}
	\hat{\tau}_l^{\text{init}}=\mathop{\mathrm{argmax}}\limits_{\tau_l }\Bigg{\{}\sum_{m=1}^{M}\left \| \textbf{Y}_{m,l}(\boldsymbol{f})e^{j2\pi \boldsymbol{f}\tau _{l}} \right \|^2\Bigg{\}}.
\end{equation} 

\indent  Then the initial estimation of AoA, $\hat{\theta}_{l}$ is calculated by
\begin{align}	 \label{doa-init}
	\hat{\theta}_{l}= \mathop{\mathrm{argmax}}\limits_{{\theta}_{l}}\left \| \textbf{C}^{H}({\theta}_{l})\textbf{Y}_{l}(\boldsymbol{f})e^{j2\pi \boldsymbol{f}\hat{\tau}_l^{\text{init}}} \right \|^2
\end{align}
where $\textbf{C}({\theta}_{l})\in \mathbb{C}^{M\times 1}$ represents the array steering vector of Rx used in the coarse search step and is calculated based on the far-field assumption. 
\subsubsection{Visibility-region-aided identification of near-field and far-field MPCs}
The difference between the near-field and far-field MPCs lies in the variation trends of the phase along the array axis, i.e., the spherical wavefront will cause the non-linear change of phase along the array axis. However, under spatial non-stationarity, an MPC may only be visible to a subset of the array elements, and the elements outside its VR contain only noise. Directly fitting the phase over the entire array aperture therefore dilutes the phase-curvature evidence with noise-only elements and may cause misclassification, especially for MPCs with short VRs. To address this issue, we propose a VR-aided weighted wideband classification method, in which the VR of each MPC is first pre-detected, and the phase fitting is then performed only within the detected VR with signal-to-noise ratio (SNR)-dependent weights, jointly over all $K$ frequency points.

a) \textit{VR pre-detection}: After the coarse delay estimation, the delay-aligned observation of the $l$-th MPC on the $k$-th frequency point and $m$-th antenna element is formed by 
\begin{align}
\label{deramp}
\tilde{Y}_{l}[m,k]=Y_{l}[m,k]\,e^{j2\pi f_k \hat{\tau}_l^{\mathrm{init}}}.
\end{align}
The delay-aligned power at the $m$-th element is calculated by
\begin{align}
\label{elepower}
p_{l,m}=\frac{1}{K}\sum_{k=1}^{K}\left|\tilde{Y}_{l}[m,k]\right|^2.
\end{align}
If the $l$-th MPC is invisible at the $m$-th element, $\tilde{Y}_{l}[m,k]$ contains only noise, and $2K p_{l,m}/\sigma_n^2$ follows a chi-square distribution with $2K$ degrees of freedom. Since $K$ is large ($K=1001$ in our system), $p_{l,m}$ is well approximated as Gaussian distributed with mean $\sigma_n^2$ and variance $\sigma_n^4/K$. Therefore, the $m$-th element is declared visible to the $l$-th MPC if
\begin{align}
\label{cfar}
p_{l,m}\geq \eta\,\hat{\sigma}_n^2,\quad \eta = 1+\frac{Q^{-1}(P_{\mathrm{fa}})}{\sqrt{K}}
\end{align}
where $Q^{-1}(\cdot)$ denotes the inverse Q-function, $P_{\mathrm{fa}}$ is the target false-alarm probability ($P_{\mathrm{fa}}=10^{-3}$ throughout this work, yielding $\eta\approx1.10$), and $\hat{\sigma}_n^2$ is estimated from the residual CTF after all $L$ MPCs are extracted. To suppress sporadic false alarms, the estimated VR of the $l$-th MPC, denoted by $\Omega_l=\{m_{l,s},\dots,m_{l,e}\}$, is taken as the longest contiguous run of visible elements, where isolated invisible gaps shorter than $G=2$ elements are filled and runs shorter than $M_{\min}=8$ elements are discarded. If no qualified VR is found, the MPC is conservatively labeled as a far-field MPC with $\Omega_l=\{1,\dots,M\}$. The cardinality of $\Omega_l$ is denoted by $M_{\mathrm{eff},l}=|\Omega_l|$.

b) \textit{Element-wise reliability weights}: Within $\Omega_l$, the reliability of the phase observation at each element depends on its post-deramping SNR, which is estimated as
\begin{align}
\label{elemsnr}
\gamma_{l,m}=\frac{\max\left(p_{l,m}-\hat{\sigma}_n^2,\,0\right)}{\hat{\sigma}_n^2},
\end{align}
since the phase-noise variance at the $m$-th element is inversely proportional to $\gamma_{l,m}$, i.e., $\mathrm{var}(\phi_l[m,k])\approx 1/(2K\gamma_{l,m})$. Each element is accordingly weighted by
\begin{align}
\label{weight}
w_{l,m}=\frac{\gamma_{l,m}}{\max\limits_{m'\in\Omega_l}\gamma_{l,m'}},\quad m\in\Omega_l,
\end{align}
so that weak elements, whose phases are dominated by noise, are automatically down-weighted rather than discarded by a hard decision.

c) \textit{Wideband joint phase model}: Within $\Omega_l$, the unwrapped phase $\phi_l[m,k]=\angle \tilde{Y}_{l}[m,k]$ of the $l$-th MPC at the $m$-th element and the $k$-th frequency point is modeled under two competing hypotheses as
\begin{align}
\label{phasemodel}
\text{Linear:}\quad &\phi_l[m,k]=\phi_0+\frac{2\pi}{\lambda_k}(m-1)b_1+\varepsilon_{m,k},\nonumber\\
\text{Quadratic:}\quad &\phi_l[m,k]=\phi_0+\frac{2\pi}{\lambda_k}\left[(m-1)b_1+(m-1)^2b_2\right]\nonumber\\&+\varepsilon_{m,k},
\end{align}
where $b_1=\delta_d\sin\theta_l$ and $b_2=\delta_d^2/(2d_{\mathrm{Rx},l})$ characterize the slope and the curvature of the phase progression, respectively, and $\varepsilon_{m,k}$ denotes the phase noise. It is worth noting that the unknown parameters $(\phi_0,b_1,b_2)$ are common to all $K$ frequency points, since the frequency dependence is entirely captured by the known factor $2\pi/\lambda_k$. Therefore, all $M_{\mathrm{eff},l}K$ phase samples are fitted jointly, instead of collapsing the CTF onto a single reference frequency, which markedly sharpens the separation between the two hypotheses at low SNR.

d) \textit{Weighted LS fitting and Bayesian classification}: Stacking the phases $\phi_l[m,k]$ ($m\in\Omega_l$, $k=1,\dots,K$) into the vector $\boldsymbol{\phi}_l\in\mathbb{R}^{M_{\mathrm{eff},l}K}$, the design matrices of the two models are given by
\begin{align}
\label{design}
\mathbf{X}_1=[\mathbf{1},\,\mathbf{u}],\quad \mathbf{X}_2=[\mathbf{1},\,\mathbf{u},\,\mathbf{u}\odot\mathbf{u}],
\end{align}
where $\mathbf{u}[m,k]=(2\pi/\lambda_k)(m-1)$. The weight matrix is $\mathbf{W}=\mathrm{diag}(\mathbf{w}_l)\otimes\mathbf{I}_K$ with $\mathbf{w}_l=[w_{l,m_{l,s}},\dots,w_{l,m_{l,e}}]^{T}$. The weighted least-squares (WLS) estimates and the corresponding weighted residual variances are obtained by
\begin{align}
\label{wls}
\hat{\boldsymbol{\beta}}_i&=(\mathbf{X}_i^{\mathsf T}\mathbf{W}\mathbf{X}_i)^{-1}\mathbf{X}_i^{\mathsf T}\mathbf{W}\boldsymbol{\phi}_l,\quad i\in\{1,2\},\\
\label{wlsvar}
\hat{\sigma}_{l,i}^2&=\frac{\left\|\mathbf{W}^{1/2}\left(\boldsymbol{\phi}_l-\mathbf{X}_i\hat{\boldsymbol{\beta}}_i\right)\right\|^2}{M_{\mathrm{eff},l}K-q_i}
\end{align}
with $q_1=2$ and $q_2=3$. According to the Bayesian information criterion (BIC), the wideband weighted BIC of each hypothesis is then computed as
\begin{align}
\label{wbic}
\mathrm{BIC}_{l,i}=M_{\mathrm{eff},l}K\left[\ln\left(2\pi\hat{\sigma}_{l,i}^2\right)+1\right]+q_i\ln\left(M_{\mathrm{eff},l}K\right).
\end{align}
If the linear model has a lower BIC, i.e., $\mathrm{BIC}_{l,1}<\mathrm{BIC}_{l,2}$, the MPC will be identified as a far-field MPC. Otherwise, the MPC will be identified as a near-field MPC. The decision margin $\Delta\mathrm{BIC}_l=\mathrm{BIC}_{l,1}-\mathrm{BIC}_{l,2}$ is retained as a per-MPC confidence measure of the classification.

\indent \textit{Remark 1 (Phase unwrapping):} Since $\delta_d=\lambda/2$, the inter-element phase step never exceeds $\pi$, so the phase $\phi_l[m,k]$ can be reliably unwrapped first along the element index $m$ and then along the frequency index $k$. The weights in (\ref{weight}) further suppress unwrapping errors caused by weak elements.

\indent \textit{Remark 2 (Synergy with birth-death identification):} The pre-detected VR boundaries $\{m_{l,s},m_{l,e}\}$ also serve as the initialization of $\hat{\xi}_{l,s}$ and $\hat{\xi}_{l,e}$ in the subsequent birth-death identification step, which confines its search range from the whole array to the neighborhood of the VR boundaries.
\subsubsection{Parameter refinement}
The refined search will be conducted at this step separately considering the spherical wavefront and planar wavefront assumptions to improve the estimation accuracy. 

When the $l$-th MPC is a near-field MPC, i.e., $l\in{\cal C}_N$, the near-field parameters $\theta_{l}$ and $d_{l}$ can be estimated by solving the following problem as
\begin{align}
\label{theta}
	\{\hat{\theta}_{l},\hat{d}_{l}\}= \mathop{\mathrm{argmax}}\limits_{\hat{\theta}_{l},\ \hat{d}_{l}}\left \| \textbf{C}_N^{H}(\hat{\theta}_{l},\hat{d}_{l})\textbf{Y}_{l}(\boldsymbol{f})e^{j2\pi \boldsymbol{f}\hat{\tau}_l^{\text{init}}} \right \|^2
\end{align}
in which $\textbf{C}_N(\hat{\theta}_{l},\hat{d}_{l})$ is the steering vector under the near-field assumption, i.e.,
\begin{align}
&\textbf{C}_N(\hat{\theta}_{l},\hat{d}_{l})=\nonumber\\&\left[  1, e^{\frac{j2\pi}{\lambda}({\delta _d}\sin {\hat{\theta}_{l}} + \frac{{{\delta _d}^2}}{{2{\hat{d}_{l}}}})}, ...,\ e^{\frac{j2\pi}{\lambda}({\delta _d}(M - 1)\sin {\hat{\theta}_{l}} + \frac{{{\delta _d}^2{{(M - 1)}^2}}}{{2{\hat{d}_{l}}}})}  \right].
\end{align}
Then, the refinement of $\tau_l$ is conducted by solving the following problem as
\begin{align}	 
\hat{\tau_l}= \mathop{\mathrm{argmax}}\limits_{\hat{\tau_l}}\left \| \textbf{C}_N^{H}(\hat{\theta}_{l},\hat{d}_{l})\textbf{Y}_{l}(\boldsymbol{f})e^{j2\pi \boldsymbol{f}\hat{\tau}_l} \right \|^2.
\end{align}
At last, the estimation of $ \hat{\boldsymbol{\alpha}}_l $ is obtained~by
\begin{equation}  
\label{alpha}
\hat{{\alpha}}_l=\frac{1}{K}\textbf{C}_N^{H}(\hat{\theta}_{l},\hat{d}_{l})\textbf{Y}_{l}(\boldsymbol{f})e^{j2\pi \boldsymbol{f}\hat{\tau}_l}.
\end{equation}  
When the $l$-th MPC is a far-field MPC, i.e., $l\in{\cal C}_F$, the far-field parameters $\theta_{l}$ and $\tau_{l}$ can be estimated by replacing the near-field steering vector $\textbf{C}_N^{H}(\hat{\theta}_{l},\hat{d}_{l})$ with the far-field steering vector $\textbf{C}_F^{H}(\hat{\theta}_{l})$ in (\ref{theta})-(\ref{alpha}).
\subsubsection{MPCs birth-death identification}
The birth-death mask matrices $\xi_{l,s}$ and $\xi_{l,e}$ are estimated alternately in this step. Instead of an exhaustive search over the whole array, the search ranges of $\xi_{l,s}$ and $\xi_{l,e}$ are confined to the neighborhood of the VR boundaries $m_{l,s}$ and $m_{l,e}$ pre-detected in the classification stage, which reduces the computational load and avoids spurious detections. Maximizing the likelihood in (\ref{Likely}) reduces to minimizing the squared Frobenius norm of the residual vector $\|\operatorname{vec}(\textbf{Y}_l)-\operatorname{vec}(\textbf{H}(\boldsymbol{f};\boldsymbol{\Gamma}_l))\|^2$. Thus, the estimation of the start antenna element index $\xi_{l,s}$  can be obtained by
\begin{align} 	\label{zeta-r}
	\hat\xi_{l,s} = & \mathop{\mathrm{argmin}}\limits_{1\leq \xi_{l,s}\leq M} \big \| \mathrm{vec} \{\textbf{Y}_{l}(\boldsymbol{f})\}
	\notag\big.
	\nonumber \\  \big.  -&\mathrm{vec}\{\textbf{H}(\boldsymbol{f};\hat{\boldsymbol{\Gamma}} _l)\}\odot\left[ {\bf{O}}_{({\xi _{l,s}-1)K}},{\bf E}_{({M}-{\xi _{l,s}}+1)K}\right ]\big \|^{2}
\end{align} 
where $\hat{\boldsymbol{\Gamma}} _l=\left \{{\hat{d}_{{\rm{Rx}},l}},{\hat\tau _l},{\hat\theta _{{\rm{Rx}},l}}\right \}$. Then, the end antenna element index $\xi_{l,e}$  can be obtained following the same principle~as
\begin{align} 	\label{zeta-r}
	\hat\xi_{l,e} = & \mathop{\mathrm{argmin}}\limits_{\hat\xi_{l,s}< \xi_{l,e}\leq M} \big \| \mathrm{vec} \{\textbf{Y}_{l}(\boldsymbol{f})\}
	\notag\big.-\mathrm{vec}\{\textbf{H}(\boldsymbol{f};\hat{\boldsymbol{\Gamma}} _l)\}
	\nonumber \\  \big.  \odot&\left[ {\bf{O}}_{({\hat\xi _{l,s}-1)K}},{\bf E}_{({{\xi _{l,e}}}-{\hat\xi _{l,s}}+1)K},{\bf{O}}_{({M-\xi _{l,e})K}}\right ]\big \|^{2}.
\end{align} 
\subsubsection{Iterative estimation}
\begin{figure}
    \centering
    \includegraphics[width=0.85\linewidth]{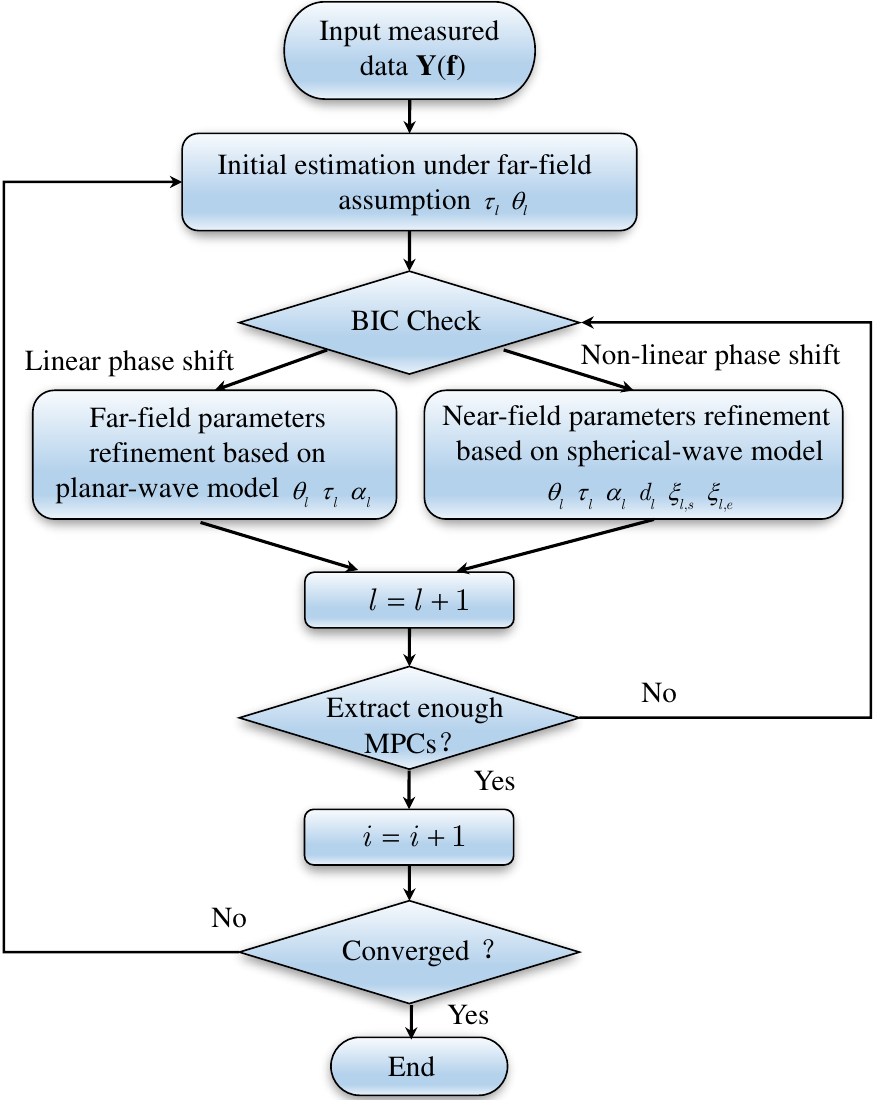}
    \caption{Flowchart of the proposed hybrid-field parameter estimation process.}
    \label{Fig_flowchart}
\end{figure}

Until now, the estimated parameter set $\hat{\boldsymbol{\Gamma}} _l$ has been obtained,  iterative refinement of $\hat{\boldsymbol{\Gamma}} _l$ is then conducted until convergence. In each iteration, the elements in the parameter set $\hat{\boldsymbol{\Gamma}} _l$ are updated in the order of AoA, distance between scatterer and Rx, delay, birth-death coefficient, and complex amplitude. The procedure proceeds as follows:

\begin{align} 	\label{doa-it}
	\left \{ \hat{\theta}_{l}^{i+1},\hat{d}_{l}^{i+1} \right \} = &\mathop{\mathrm{argmax}}\limits_{{\theta}_{l}^{i+1},{d}_{l}^{i+1}}\big\{z\big(\boldsymbol{f} ; {\hat{\tau _l}^{i},{\hat\xi _{l,s}^{i}},{\hat\xi _{l,e}^{i}}} \big)\big\}
\end{align} 
\begin{align} 	\label{doa-it}
	\hat{\tau}_{l}^{i+1} = &\mathop{\mathrm{argmax}}\limits_{{\tau}_{l}^{i+1}}\big\{z\big(\boldsymbol{f} ; \hat{\theta}_{l}^{i+1},\hat{d}_{l}^{i+1} ,{\hat\xi _{l,s}^{i}},{\hat\xi _{l,e}^{i}} \big)\big\}
\end{align} 
\begin{align} 	\label{doa-it}
	 &\left\{ {\hat\xi _{l,s}^{i+1}},{\hat\xi _{l,e}^{i+1}} \right\}= 
	 \mathop{\mathrm{argmin}}\limits_{{\xi _{l,s}^{i+1}},{\xi _{l,e}^{i+1}}} \big \| \mathrm{vec} \{\textbf{Y}_{l}(\boldsymbol{f})\}
	\notag\big.-\mathrm{vec}\{\textbf{H}(\boldsymbol{f};\hat{\boldsymbol{\Gamma}} _l^{i+1})\}
	\nonumber \\  \big.  &\odot\left[ {\bf{O}}_{({\xi^{i+1} _{l,s}-1)K}},{\bf E}_{({{\xi^{i+1} _{l,e}}}-{\xi^{i+1} _{l,s}}+1)K},{\bf{O}}_{({M-\xi ^{i+1}_{l,e})K}}\right ]\big \|^{2}
\end{align} 
\begin{equation}  
\hat{{\alpha}}_l^{i+1}=\frac{1}{K} {\bf{\hat\xi }}_{l}^{i+1}\odot\textbf{C}_N^{H}(\hat{\theta}_{l}^{i+1},\hat{d}_{l}^{i+1})\textbf{Y}_{l}(\boldsymbol{f})e^{j2\pi \boldsymbol{f}\hat{\tau}_l^{i+1}}
\end{equation} 
where $\hat{[\cdot]}^{i}$ denotes the $i$-th iteration estimation results of the given argument, $z{(\cdot)}$ denotes the objective function defined in \cite{ZhouSAGE}, ${\bf{\hat\xi }}_{l}^{i+1}$ is generated by substituting ${\hat\xi _{l,s}^{i+1}}$ and ${\hat\xi _{l,e}^{i+1}}$ into (\ref{Mask}), and the first estimated $\hat{\boldsymbol{\Gamma}} _l$ is expressed as~ $\hat{\boldsymbol{\Gamma}} _l^{1}$. At the end of each iteration cycle, the value of the likelihood function in (\ref{Likely}) is computed and compared with the likelihood value in the last iteration. If the increase of the likelihood function compared with that of the last iteration is smaller than a certain threshold, the iteration is judged to have converged.
\begin{align}
    {\cal L}({\bf{\Gamma }}^{i+1};{\bf{Y}}({\boldsymbol{f}}))-{\cal L}({\bf{\Gamma }}^{i};{\bf{Y}}({\boldsymbol{f}}))<r{\cal L}({\bf{\Gamma }}^{i};{\bf{Y}}({\boldsymbol{f}})).
\end{align}
Here, $r$ denotes the relative likelihood improvement threshold. Iteration also stops once the cycle count surpasses a certain number, e.g. 10 in this study. For clarity, the hybrid-field parameter estimation process is illustrated in Fig. \ref{Fig_flowchart}.

The complexity of the proposed hybrid-field SAGE algorithm is
$\mathcal{O}(N_{\mathrm{g}}MK)$ per iteration, where $N_{\mathrm{g}}$ is the total number of grid candidates. Compared with the purely near-field SAGE scheme \cite{ZhouSAGE}, the additional VR pre-detection and weighted BIC classification only involve closed-form operations with $\mathcal{O}(MK)$ cost, which is negligible against the grid searches. In return, the classification reduces the overall search overhead: an MPC classified as far field only requires a one-dimensional angle-only refinement instead of a two-dimensional angle-distance search. Therefore, the proposed scheme has less complexity than the purely near-field scheme, and its complexity further decreases as the proportion of far-field MPCs increases. 

\subsection{Validation with Synthetic Data}
To validate the proposed VR-aided classification under controlled conditions with known ground truth, Monte Carlo simulations are conducted on synthetic hybrid-field channels generated according to (2). The array and sounding parameters follow the measurement system, i.e., $M=128$ with $\delta_d=\lambda/2$, and a 20 GHz bandwidth with $K=1001$ frequency points at 300 GHz. Two scenarios with $L=6$ and $L=8$ MPCs are considered, whose ground-truth parameters are listed in Tables~\ref{Tab_sim_scene_6} and \ref{Tab_sim_scene_8}, respectively. In both scenarios, near-field (NF) MPCs with spherical wavefronts, limited VRs, and non-uniform power profiles coexist with far-field (FF) MPCs with planar wavefronts and full-aperture visibility. Compared with the 6-path scenario, the 8-path scenario features denser multipath, a larger power dynamic range, and a wider angular spread, which further stresses the estimation in dense multipath conditions. The complex gain of each MPC is given by $\alpha_l=|\alpha_l|e^{j\phi_l}$, where $\phi_l$ is independently drawn from a uniform distribution over $[0,2\pi)$ in each trial. Note that the synthetic channels are generated with the exact spherical wavefront, while the estimator adopts the second-order Taylor-expanded model in (2), which intentionally introduces a model mismatch as encountered in practice. For each SNR, 200 independent trials are performed, where the SNR is defined with respect to the reference MPC~as
\begin{align}
\label{snr_def}
{SNR}&=\frac{P_{\mathrm{ref}}}{\sigma_n^2}\\
P_{\mathrm{ref}}=\frac{1}{K\,M_{\mathrm{VR},1}}&\sum_{m\in\Omega_1}\sum_{k=1}^{K}\left|H_1[m,k]\right|^2
\end{align}
where $H_1[m,k]$ denotes the CTF of the reference MPC and $M_{\mathrm{VR},1}=|\Omega_1|$ is the length of its VR.

\begin{table}[!t]
\centering
\caption{Ground-Truth Parameters of the 6-Path Synthetic Scenario}
\label{Tab_sim_scene_6}
\begin{tabular}{cccccc}
\hline
MPC & $|\alpha_l|$ & $\tau_l$ (ns) & $\theta_l$ ($^\circ$) & Type ($d_{\mathrm{Rx},l}$, m) & VR $[\xi_{l,s},\xi_{l,e}]$\\
\hline
1 & 1.00 & 9.63  & 1.5    & NF (0.8) & [8, 120]\\
2 & 0.95 & 5.13  & $-$2.5 & NF (1.0) & [18, 110]\\
3 & 0.91 & 14.13 & 7.5    & FF & [1, 128]\\
4 & 0.87 & 18.63 & $-$4.5 & NF (1.3) & [12, 100]\\
5 & 0.82 & 23.13 & 5.5    & FF & [1, 128]\\
6 & 0.78 & 27.63 & $-$6.5 & FF & [1, 128]\\
\hline
\end{tabular}
\end{table}

\begin{table}[!t]
\centering
\caption{Ground-Truth Parameters of the 8-Path Synthetic Scenario}
\label{Tab_sim_scene_8}
\begin{tabular}{cccccc}
\hline
MPC & $|\alpha_l|$ & $\tau_l$ (ns) & $\theta_l$ ($^\circ$) & Type ($d_{\mathrm{Rx},l}$, m) & VR $[\xi_{l,s},\xi_{l,e}]$\\
\hline
1 & 1.00 & 8.13  & 1.5    & NF (0.9) & [5, 120]\\
2 & 0.93 & 13.63 & $-$2.5 & FF & [1, 128]\\
3 & 0.86 & 19.13 & 4.0    & NF (1.4) & [20, 90]\\
4 & 0.79 & 24.63 & $-$5.5 & FF & [1, 128]\\
5 & 0.72 & 30.13 & 7.0    & NF (2.0) & [35, 105]\\
6 & 0.65 & 35.63 & $-$8.5 & FF & [1, 128]\\
7 & 0.58 & 41.13 & 10.0   & NF (4.8) & [50, 120]\\
8 & 0.48 & 46.63 & $-$11.5 & FF & [1, 128]\\
\hline
\end{tabular}
\end{table}

\begin{figure}[!t]
    \centering
    \includegraphics[width=0.80\linewidth]{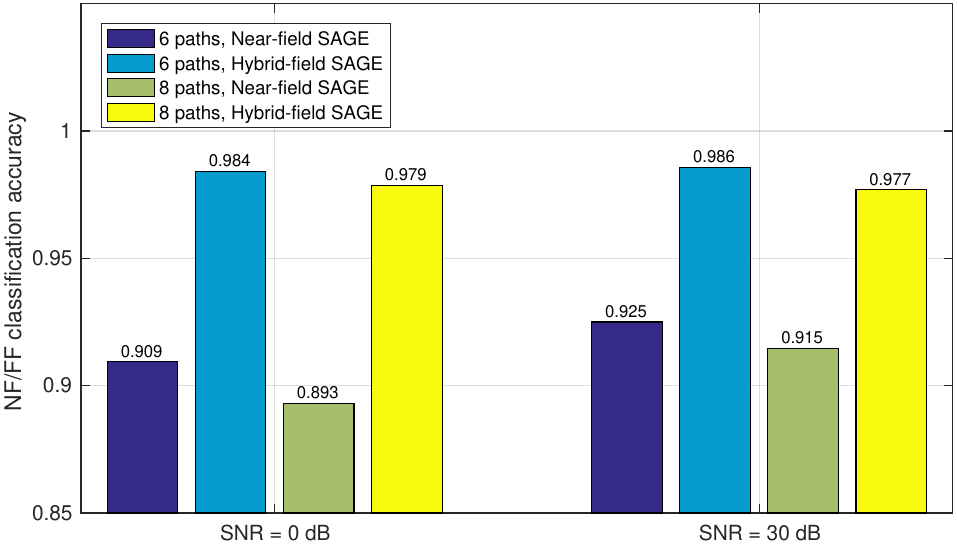}
    \caption{NF/FF classification accuracy of the near-field SAGE algorithm \cite{ZhouSAGE} and the proposed hybrid-field SAGE algorithm at SNR = 0 dB and 30~dB.}
    \label{Fig_sim_acc}
\end{figure}

Fig.~\ref{Fig_sim_acc} compares the NF/FF classification accuracy of the near-field SAGE classifier \cite{ZhouSAGE} and the proposed VR-aided hybrid-field SAGE classifier at SNR = 0 dB and 30 dB. For the near-field SAGE \cite{ZhouSAGE}, the NF/FF decision is obtained according to the distance from scatterers to Rx compared with Rayleigh distance. The proposed classifier improves the accuracy from 0.91 to 0.98 ($L=6$) and from 0.89 to 0.98 ($L=8$) at 0 dB, and the gain persists at 30 dB. The remaining classification errors are found to concentrate on MPCs whose scatterer distances are close to the Rayleigh distance, where spherical and planar wavefronts are barely distinguishable.

\begin{figure}[!t]
    \centering
    \includegraphics[width=0.80\linewidth]{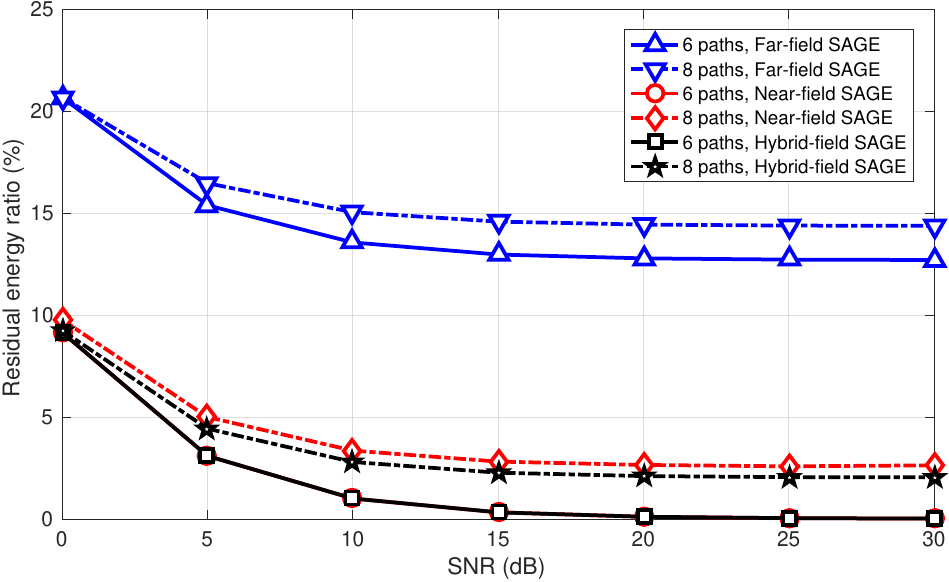}
    \caption{Residual energy ratio versus SNR for the 6-path and 8-path synthetic hybrid-field scenarios.}
    \label{Fig_sim_res}
\end{figure}

Fig.~\ref{Fig_sim_res} shows the residual energy ratio after extracting all MPCs, which is defined as
\begin{align}
\label{res_def}
P_{\mathrm{res}}=\frac{\left\|\mathbf{Y}(\boldsymbol{f})-\sum_{l=1}^{L}\mathbf{H}(\boldsymbol{f};\hat{\boldsymbol{\Gamma}}_l)\right\|_F^2}{\left\|\mathbf{Y}(\boldsymbol{f})\right\|_F^2}.
\end{align}

Specifically, three schemes are compared in Fig.~\ref{Fig_sim_res}, i.e., the far-field SAGE \cite{Xue2003} and near-field SAGE \cite{ZhouSAGE}, where all MPCs are modeled with planar and spherical wavefronts, respectively, and the proposed hybrid-field SAGE, where each MPC is assigned a wavefront model according to the proposed classifier. The far-field SAGE exhibits an SNR-insensitive residual floor of 12\%--15\%, since planar wavefronts cannot represent the spherical phase progression of near-field MPCs. In contrast, the near-field and hybrid-field schemes achieve comparable and much lower residuals, both decreasing to around 3\% or below at high SNR, with the proposed scheme attaining the lowest residual over the entire SNR range in both~scenarios. 

\section{Channel Characterization Using Real Measurement Data}
\label{Channel Characterization and Evaluation Using Real Measurement Data}
Using the estimated MPC parameters extracted from measurement data described in  Section \ref{THz Channel Measurements}, we proceed to compute several statistical properties that are commonly used to characterize propagation channels. In the data processing, MPCs are extracted until the
residual peak drops below the larger of 45 dB under the strongest MPC and 10 dB above the noise floor. The delay, angle, and distance searches use grid spacings of 25 ps, 0.14 degrees, and 1.7 cm, respectively. 
\subsection{Time Delay PSD}
\begin{figure}
    \centering
    \includegraphics[width=0.99\linewidth]{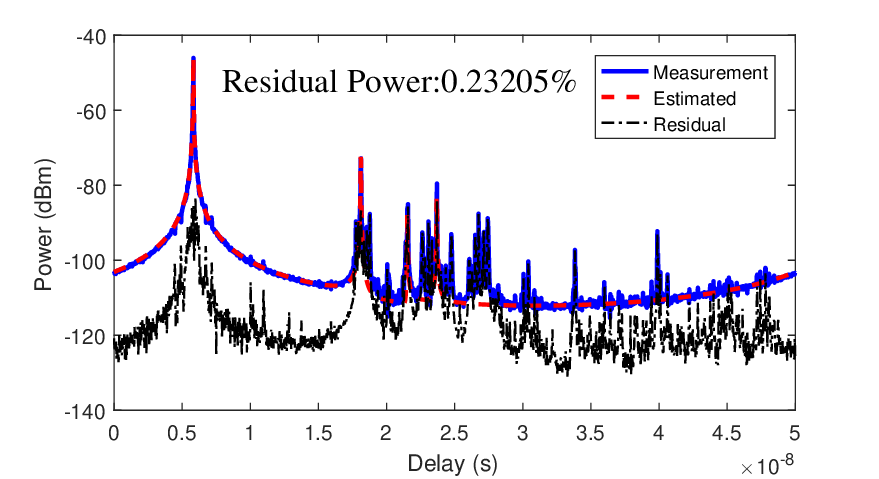}
    \caption{Delay PSDs of measurement, estimated, and residual channels (Tx position 2).}
    \label{Fig_estPDP}
\end{figure}

In Fig. \ref{Fig_estPDP}, the time delay PSD is presented. The time delay PSD shows the power distribution of the multipath channel in the time delay domain. It can be output by the VNA directly through Fourier transformation of the measurement channel transfer function. In Fig. \ref{Fig_estPDP}, we illustrate the time delay PSD outputted by the VNA, PSD generated by parameters estimated through raw data, and residual time delay PSD computed by the difference between measurement PSD and estimated PSD. It could be found that LoS path holds the majority of power in the THz channel and there are a number of distinguishable MPCs along the time delay axis and the proposed parameter estimation algorithm can extract most MPCs accurately.

\begin{figure}
    \centering
    \includegraphics[width=0.99\linewidth]{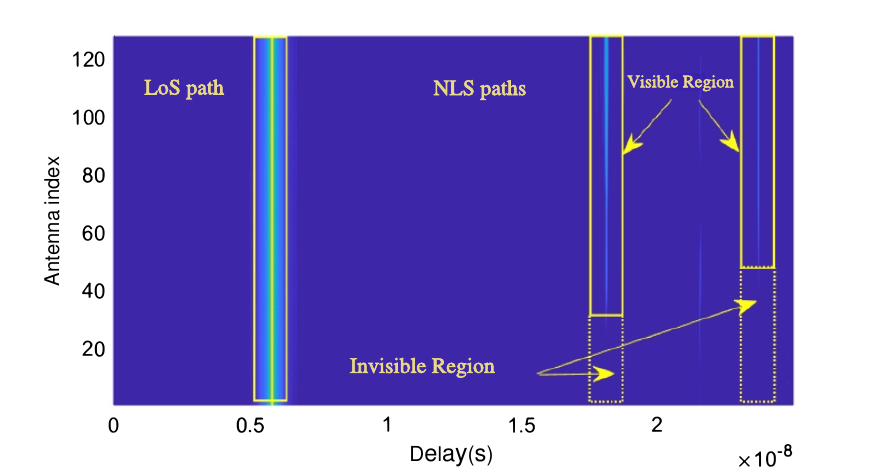}
    \caption{Illustration of visible region and invisible region (Tx position 2).}
    \label{Fig_CVR}
\end{figure}

\begin{figure}[t]
    \centering
    \includegraphics[width=0.99\linewidth]{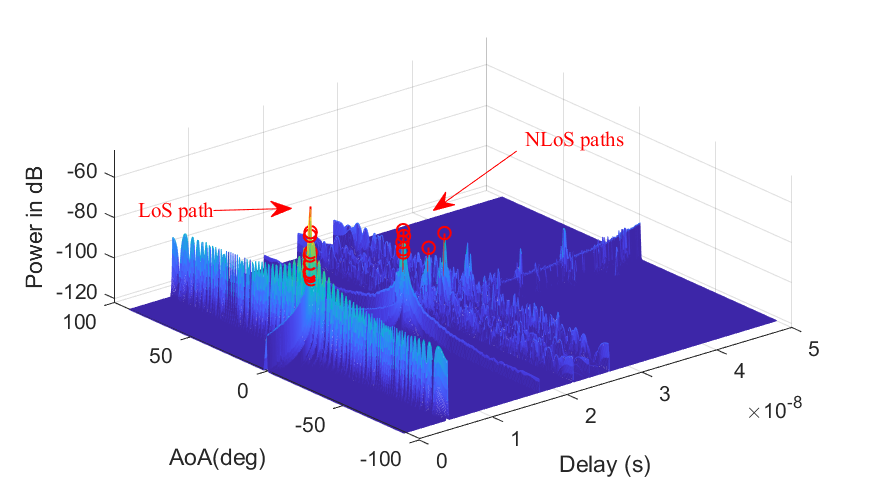}
    \caption{Delay-angle domain PSD and extracted MPCs (red dots,  Tx position 2).}
    \label{Fig_estMPC}
\end{figure}

Besides, we present the time delay PSDs for different antenna elements in Fig. \ref{Fig_CVR}. Results show that the power of LoS path maintains a constant value along the array axis. However, the NLoS paths present the birth-death phenomenon along the array axis. NLoS MPCs are only observable in a certain array window, i.e., the VR. In the context of antenna array signal processing, the VR refers to the subset of antenna elements that effectively receive a specific MPC, while the invisible region encompasses the elements where the MPC is either too weak or not detectable at all.
Based on the antenna domain delay PSD, the delay-angle domain PSD can be obtained by Fourier transform from antenna domain to angle domain. As shown in Fig. \ref{Fig_estMPC}, the delay-angle domain PSD of measurement data at 260 GHz frequency band presents several strong peaks in both delay domain and angle domain, which presents the sparsity of the THz channel in delay and angle domains. The MPCs powers and positions in delay and angle domains are also provided according to the proposed hybrid-field SAGE algorithm, i.e., the red color circles. The measurement result and estimated result match very well, illustrating the accuracy of the proposed channel estimation scheme.

\begin{figure}
    \centering
    \includegraphics[width=1.03\linewidth]{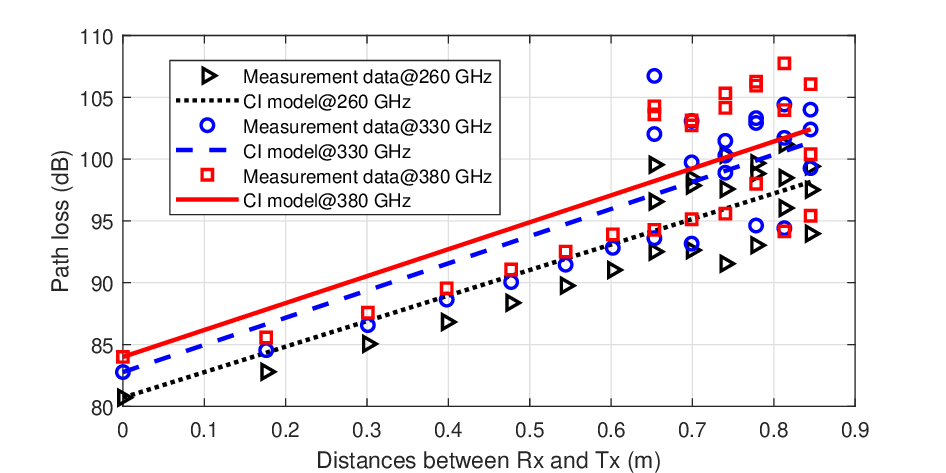}
    \caption{Measurement and CI model fitting results of path loss of indoor THz channel.}
    \label{Fig_Pathloss}
\end{figure}

\subsection{Path Loss}
The path loss analysis is illustrated in Fig. \ref{Fig_Pathloss}. Path loss is used to characterize the effect of signal attenuation caused by the increasing distance between Tx and Rx. We utilize the close-in (CI) path loss model for measurement data in this work. In general, the CI path loss model is represented as
\begin{align}
    PL = 10 \cdot n_{f_c}\cdot\log10(d)+{\rm FSPL}(d_0,f)+X_{\sigma}
\end{align}
in which $n_{f_c}$ denotes the path loss exponent, $d$ refers to the distance between the Tx and the Rx, $d_0$ refers to the reference distance between the Tx and the Rx which is set to 1 m, $f_c$ refers to the carrier frequency, and
$X_\sigma$ is the shadow fading in dB following the zero-mean Gaussian distribution. In Fig. \ref{Fig_Pathloss}, the measurement path loss results are obtained from the difference of transmitting power and receiving power. We found that the path loss exponent increases with the increase of carrier frequency at THz band. According to the fitting results in Fig. \ref{Fig_Pathloss}, the path loss exponent in measurement indoor scenario is obtained as $n_{f_c} = 2.066$ @260 GHz, $n_{f_c} = 2.098$ @330 GHz, and $n_{f_c} = 2.177$ @380 GHz,~respectively.

\subsection{Residual Power and Percentage of Near-field MPC}
\begin{figure}
    \centering
    \includegraphics[width=1.0\linewidth]{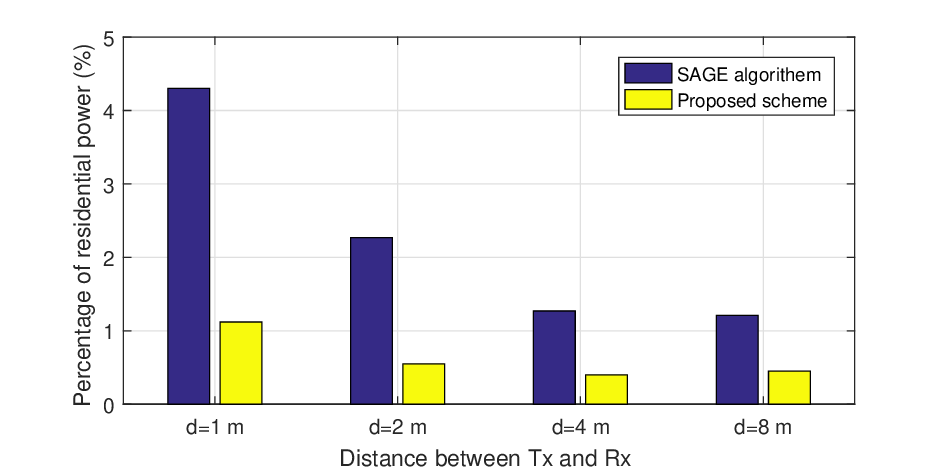}
    \caption{Residual power of the proposed method and the original SAGE method at different Tx-Rx distances.}
    \label{Fig_residential}
\end{figure}

\begin{figure}
    \centering
    \includegraphics[width=1.0\linewidth]{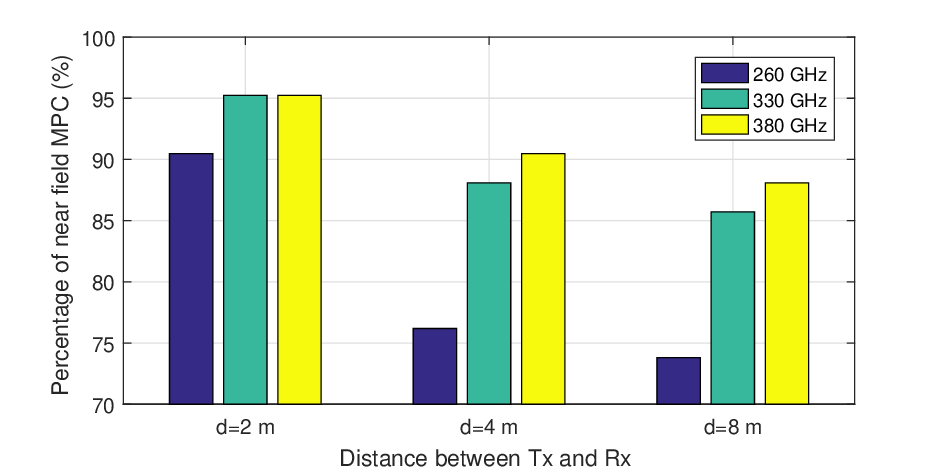}
    \caption{Percentage of near-field MPCs at different frequency bands and distances.}
    \label{Fig_NFPer}
\end{figure}

The residual power and percentage of near-field MPC of the THz measurement data are shown in Fig. \ref{Fig_residential} and Fig. \ref{Fig_NFPer}. Fig. \ref{Fig_residential} shows the residual power comparison between the proposed improved algorithm and the original SAGE algorithm at different communication distances. The proposed algorithm consistently achieves lower residual power than the original SAGE across all distances (2 m, 4 m, 8 m), highlighting its superior accuracy in capturing characteristics of the THz channel with large-scale antenna arrays like near-field effects and spatial non-stationarity. This advantage is particularly significant in short-range scenarios, crucial for enhancing channel modeling accuracy in indoor environments.

The percentage of near-field MPC quantifies the fraction of resolvable near-field MPCs in all MPCs, and is defined as 
\begin{align}
\zeta_{\mathrm{NF}} = \frac{N_{\mathrm{NF}}}{N_{\mathrm{NF}} + N_{\mathrm{FF}}} \times 100\%  
\end{align}
where \(N_{\mathrm{NF}}\) denotes the number of MPCs identified as near-field paths and \(N_{\mathrm{FF}}\) represents the far-field MPC count. Classification is performed using the proposed hybrid-field SAGE algorithm. This metric objectively characterizes the significance of near-field wave effects within the channel's multipath structure. Fig. \ref{Fig_NFPer} presents the percentage of near-field MPCs across different frequency bands (260 GHz, 330 GHz, 380 GHz) and Tx-Rx distances (2 m, 4 m, 8 m). The percentage increases with higher frequencies and shorter distances. This is because higher-frequency signals have shorter wavelengths, making the near-field effect more pronounced. At shorter distances, the spherical wavefront characteristics are more evident, leading to a higher proportion of near-field MPCs. The results indicate that in indoor scenarios, the near-field effect is significant, especially at higher frequencies and shorter distances. 

\subsection{Invisible Region and Position of Scatterer}
\begin{figure}
    \centering
    \includegraphics[width=1.05\linewidth]{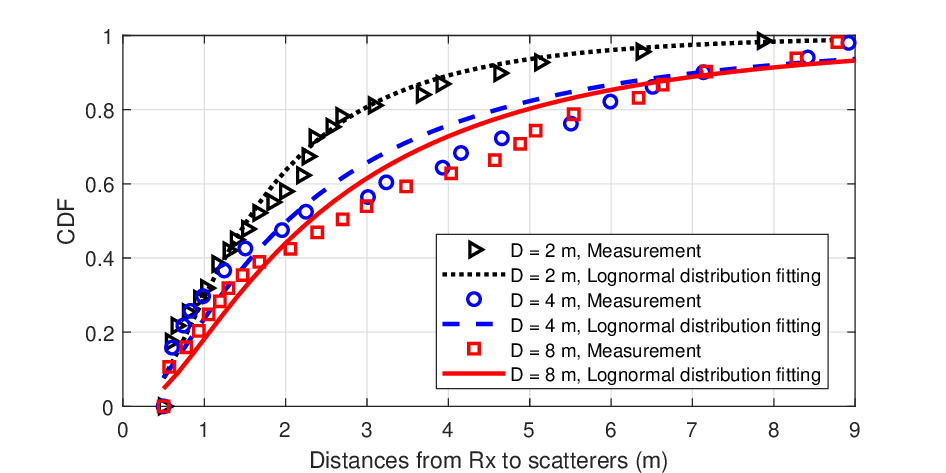}
    \caption{Distances from Rx to scatterers at different distances between Tx and Rx ($f_c = 260 $ GHz).}
    \label{Fig_Dis}
\end{figure}

\begin{figure}
    \centering
    \includegraphics[width=1.05\linewidth]{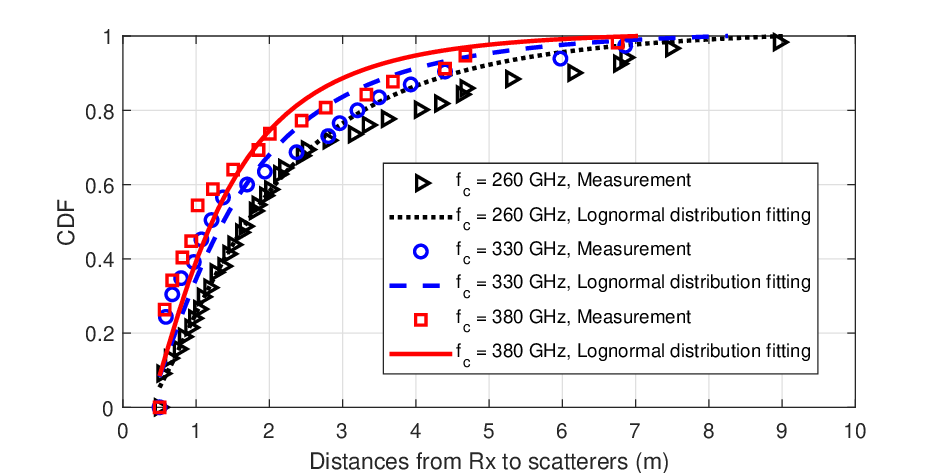}
    \caption{Distances from Rx to scatterers at different frequency bands (Tx-Rx distance $D$ = 4 m).}
    \label{Fig_Dis_f}
\end{figure}

Fig. \ref{Fig_Dis} and Fig. \ref{Fig_Dis_f} illustrate the cumulative distribution function (CDF) of the distances from the Rx to scatterers at different Tx-Rx distances and frequency bands, respectively. At a Tx-Rx distance of 2 m, the estimated distances from the Rx to scatterers are mostly within 5 m, with a few exceeding 5 m. As the Tx-Rx distance increases to 4 m and 8 m, the distribution of scatterer distances shifts slightly to longer ranges, but most scatterers are still within 5 m of the Rx. This indicates that in indoor environments, scatterers are predominantly located close to the Rx, which is typical for indoor office settings with limited space. Across different frequency bands (260 GHz, 330 GHz, 380 GHz), the CDF curves show a similar trend, with most scatterers located within 5 m of the Rx. The lognormal distribution fitting matches the measurement data well, suggesting that the distances from the Rx to scatterers can be effectively modeled using a lognormal distribution. The lognormal distribution parameters can be found in Table \ref{lognormal}.

\begin{figure}
    \centering
    \includegraphics[width=1.05\linewidth]{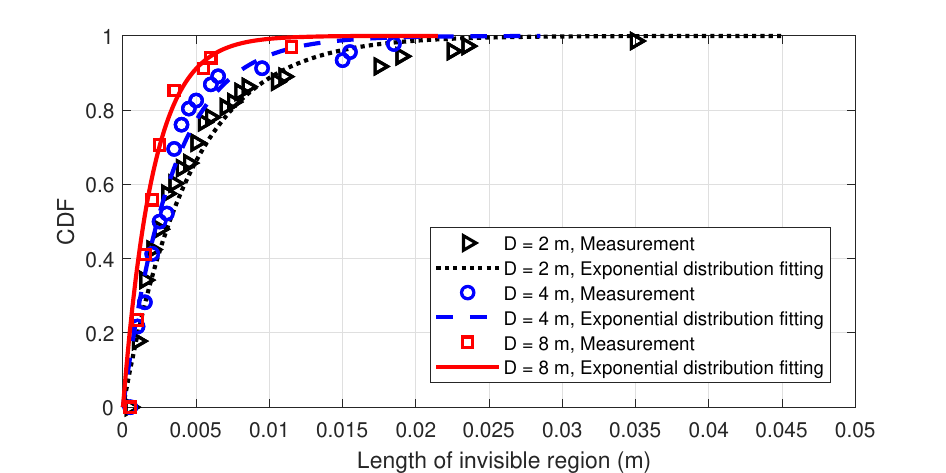}
    \caption{Lengths of invisible regions at different distances between Tx and Rx ($f_c = 260 $ GHz).}
    \label{Fig_VR}
\end{figure}

\begin{figure}
    \centering
    \includegraphics[width=1.05\linewidth]{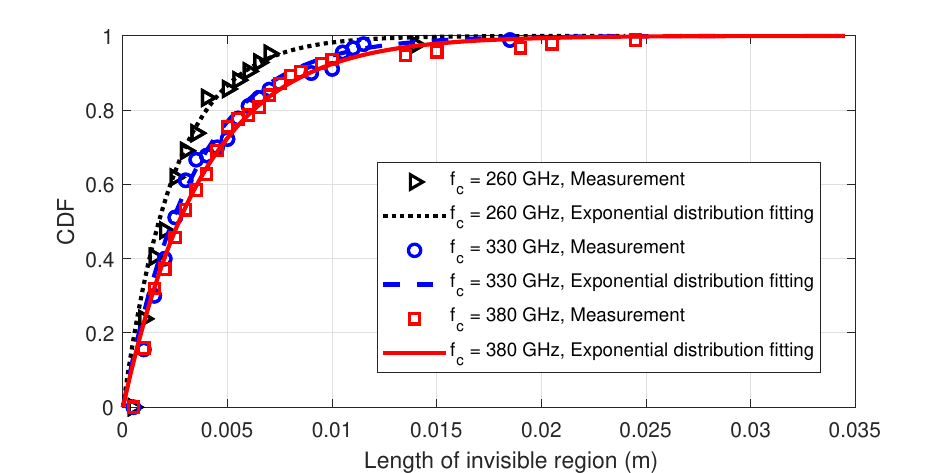}
    \caption{Lengths of invisible regions at different frequency bands (Tx-Rx distance $D$ = 4 m).}
    \label{Fig_VR_f}
\end{figure}

Fig. \ref{Fig_VR} and Fig. \ref{Fig_VR_f}  depict the CDF of the lengths of invisible regions at different Tx-Rx distances and frequency bands, respectively. The lengths of invisible regions is calculated according to estimated visible region parameters~as 
\begin{align}
    L_{{\rm CVR},l}=(M-{\hat\xi _{l,e}}+{\hat\xi _{l,s}}-1)\delta_d .
\end{align}
At a Tx-Rx distance of 2 m, the length of the invisible region is relatively small, mostly below 0.02 m. As the Tx-Rx distance increases to 4 m and 8 m, the length of the invisible region also increases slightly. This indicates that the VR of MPCs becomes slightly larger as the Tx-Rx distance increases. At different frequency bands, the length of the invisible region shows a similar trend. With the increase of carrier frequency, the length of the invisible region tends to increase. The exponential distribution fitting aligns well with the measurement data, suggesting that the length of the invisible region can be effectively modeled using an exponential distribution. The exponential distribution parameters can be found in Table \ref{exponential}.

\begin{figure}[t]
    \centering
    \includegraphics[width=1.05\linewidth]{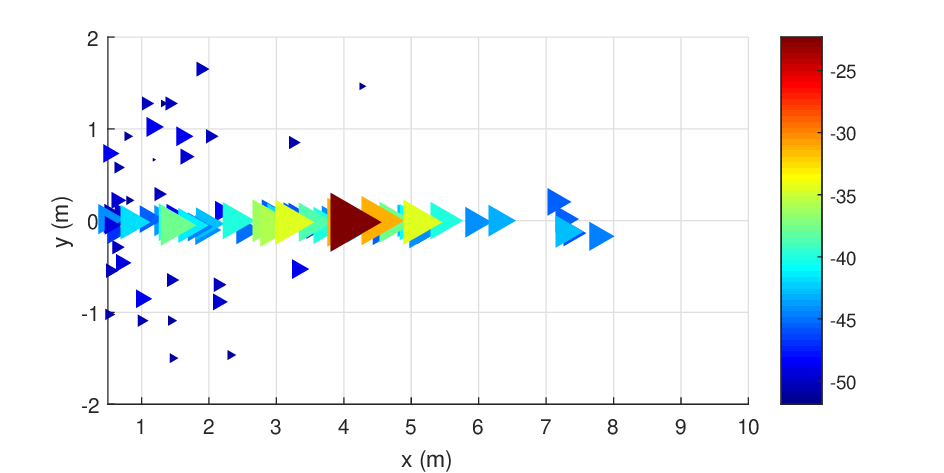}
    \caption{Scatterer distribution in 2D plane ($f_c = 260 $ GHz and Tx-Rx distance $D$ = 4 m).}
    \label{Fig_Scatterer}
\end{figure}

\begin{figure}
    \centering
    \includegraphics[width=1.05\linewidth]{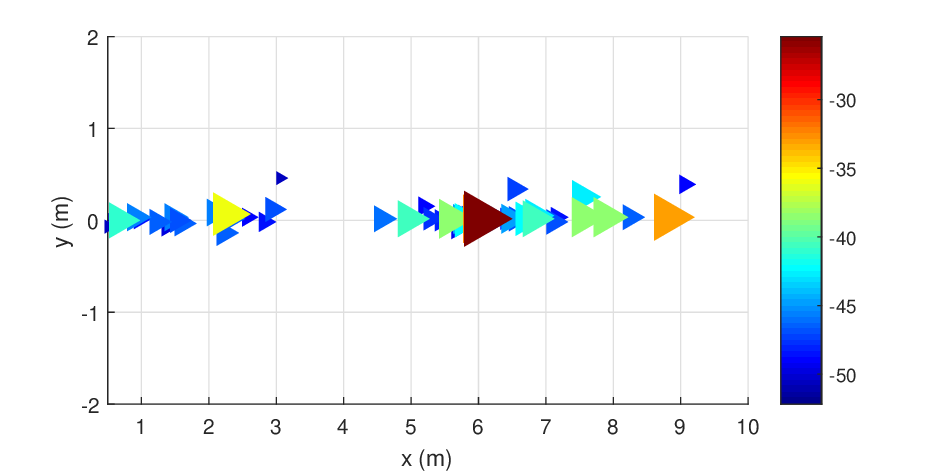}
    \caption{Scatterer distribution in 2D plane ($f_c = 260 $ GHz and Tx-Rx distance $D$ = 6 m).}
    \label{Fig_Scatterer2}
\end{figure}

Fig. \ref{Fig_Scatterer} and Fig. \ref{Fig_Scatterer2} illustrate the scatterer distribution in the two-dimensional (2D) plane. They reveal that the strongest MPCs, which align with the communication distances of 4 m and 6 m, are caused by LoS propagation from Tx to Rx. This indicates that the proposed parameter estimation algorithm can effectively extract the LoS path and other MPCs. By accurately estimating the parameters of the LoS path and other MPCs, such as their distances and angles, the algorithm enables the localization of the Tx, Rx, and scatterers in the hybrid-field indoor environment. This capability is crucial for understanding the signal propagation characteristics and optimizing the system design in indoor THz communication scenarios. The location of scatterers and Rx can also help in the sensing of propagation~environments.

\begin{table}[b]
\renewcommand{\arraystretch}{1.3}
\caption{Lognormal Distribution Parameters of Scatterer-to-Rx Distance $d_{l}$}
\label{tab:scatterer}
\centering
\begin{tabular}{c c c c}
\toprule
\textbf{Tx-–Rx} & \textbf{260 GHz} & \textbf{330 GHz} & \textbf{380 GHz} \\
\cmidrule(lr){2-4}
\textbf{Distance} & \textbf{Mean $\mu$ / Std $\sigma$} & \textbf{Mean $\mu$ / Std $\sigma$} & \textbf{Mean $\mu$ / Std $\sigma$} \\
\midrule
2 m & 0.55 / 0.75 & 0.42 / 0.73 & 0.37 / 0.79 \\
4 m & 1.01 / 0.96 & 1.04 / 0.88 & 1.04 / 0.85 \\
8 m & 1.04 / 0.74 & 1.09 / 0.89 & 1.71 / 0.89 \\
\bottomrule
\end{tabular}
\label{lognormal}
\end{table}

\begin{table}[!b]
\renewcommand{\arraystretch}{1.3}
\caption{Exponential Distribution Parameter $\lambda$ of inVisible Region $L_{\text{CVR},l}$}
\label{tab:cvr}
\centering
\begin{tabular}{c c c c}
\toprule
\textbf{Tx-–Rx} & \textbf{260 GHz} & \textbf{330 GHz} & \textbf{380 GHz} \\
\cmidrule(lr){2-4}
\textbf{Distance} & \textbf{Mean = 1/$\lambda$} & \textbf{Mean = 1/$\lambda$} & \textbf{Mean = 1/$\lambda$} \\
\midrule
2 m & 0.0041 & 0.0033 & 0.0040 \\
4 m & 0.0043 & 0.0039 & 0.0026 \\
8 m & 0.0045 & 0.0040 & 0.0038 \\
\bottomrule
\end{tabular}
\label{exponential}
\end{table}

\subsection{Space-frequency Correlation Function}
To quantify how the channel correlation behaves across the ultra-wide THz band and large-scale antenna array \cite{Wang2021}, we compute the space-frequency correlation function according to~\cite{Wang2023b,Wang2025,Wang2026}
\begin{align}
    \rho(\Delta d,\Delta f)=\mathbb{E}\!\left[H_{m}(f)\,H_{m+\Delta d}^{*}(f+\Delta f)\right]
\end{align}
where $H_{m}(f)$ is the measured CTF for each Tx--Rx horn pair, $\Delta d$ refers to the space difference across the antenna array, and $\Delta f$ refers to the frequency difference that spans the 20 GHz sweeping~bandwidth. 
\begin{figure}[t]
    \centering
    \includegraphics[width=0.87\linewidth]{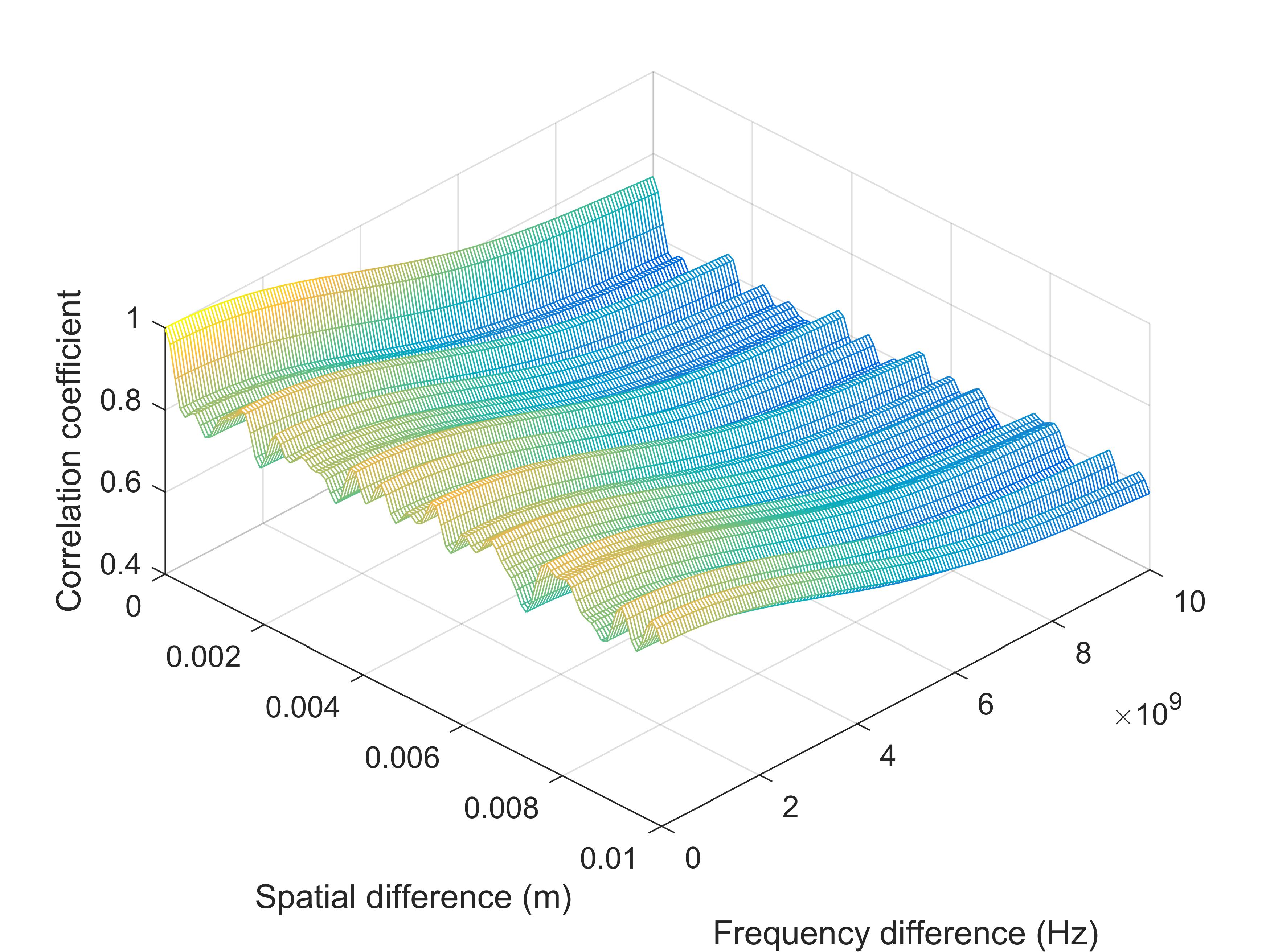}
    \caption{Space-frequency correlation functions ($f_c = 260 $ GHz and Tx-Rx distance $D$ = 3 m, Tx position 7).}
    \label{Fig_CF7}
\end{figure}
\begin{figure}
    \centering
    \includegraphics[width=0.88\linewidth]{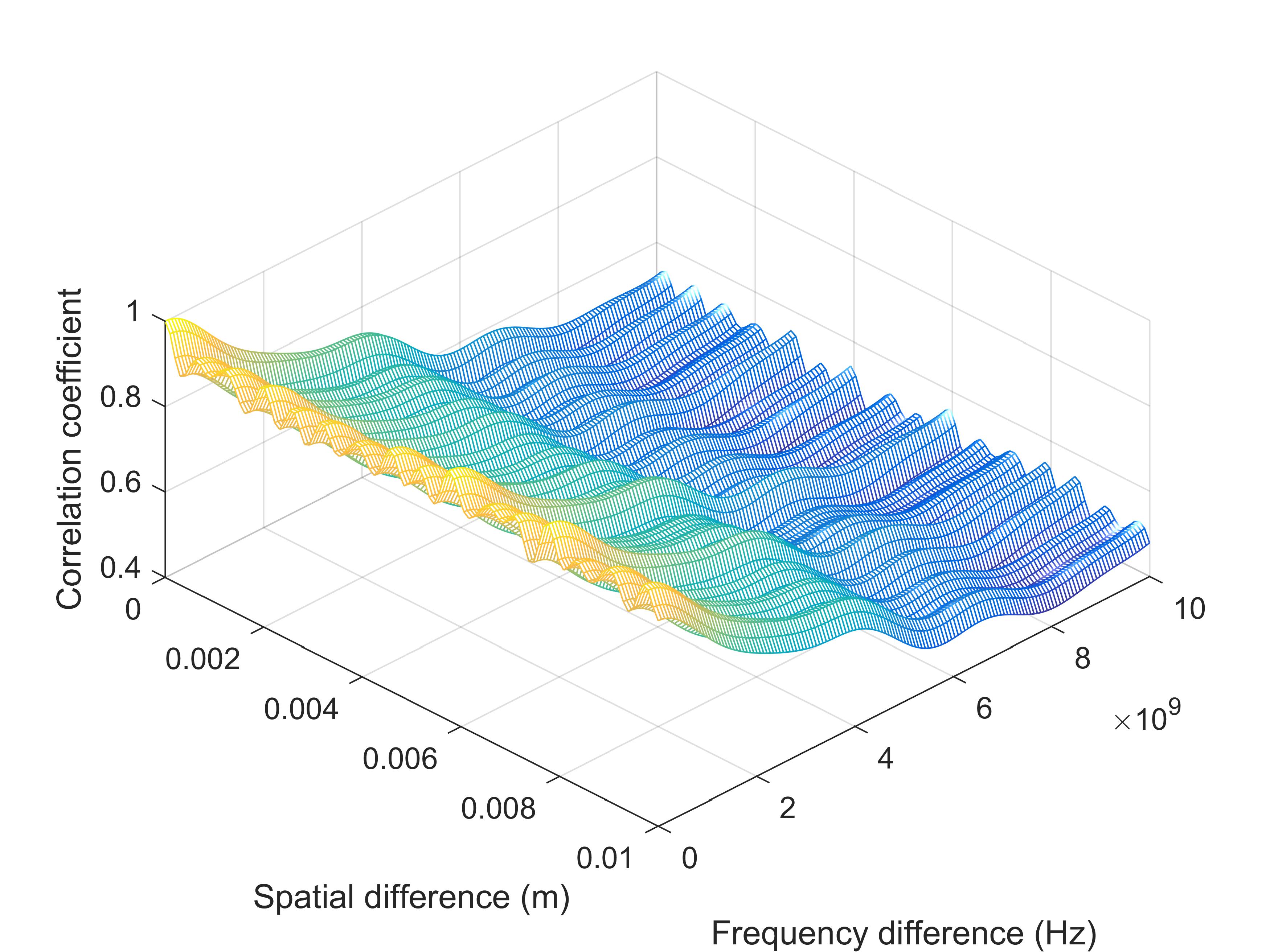}
    \caption{Space-frequency correlation functions ($f_c = 260 $ GHz and Tx-Rx distance $D$ =  6 m, Tx position 15).}
    \label{Fig_CF15}
\end{figure}
\begin{figure}[t]
    \centering
    \includegraphics[width=0.95\linewidth]{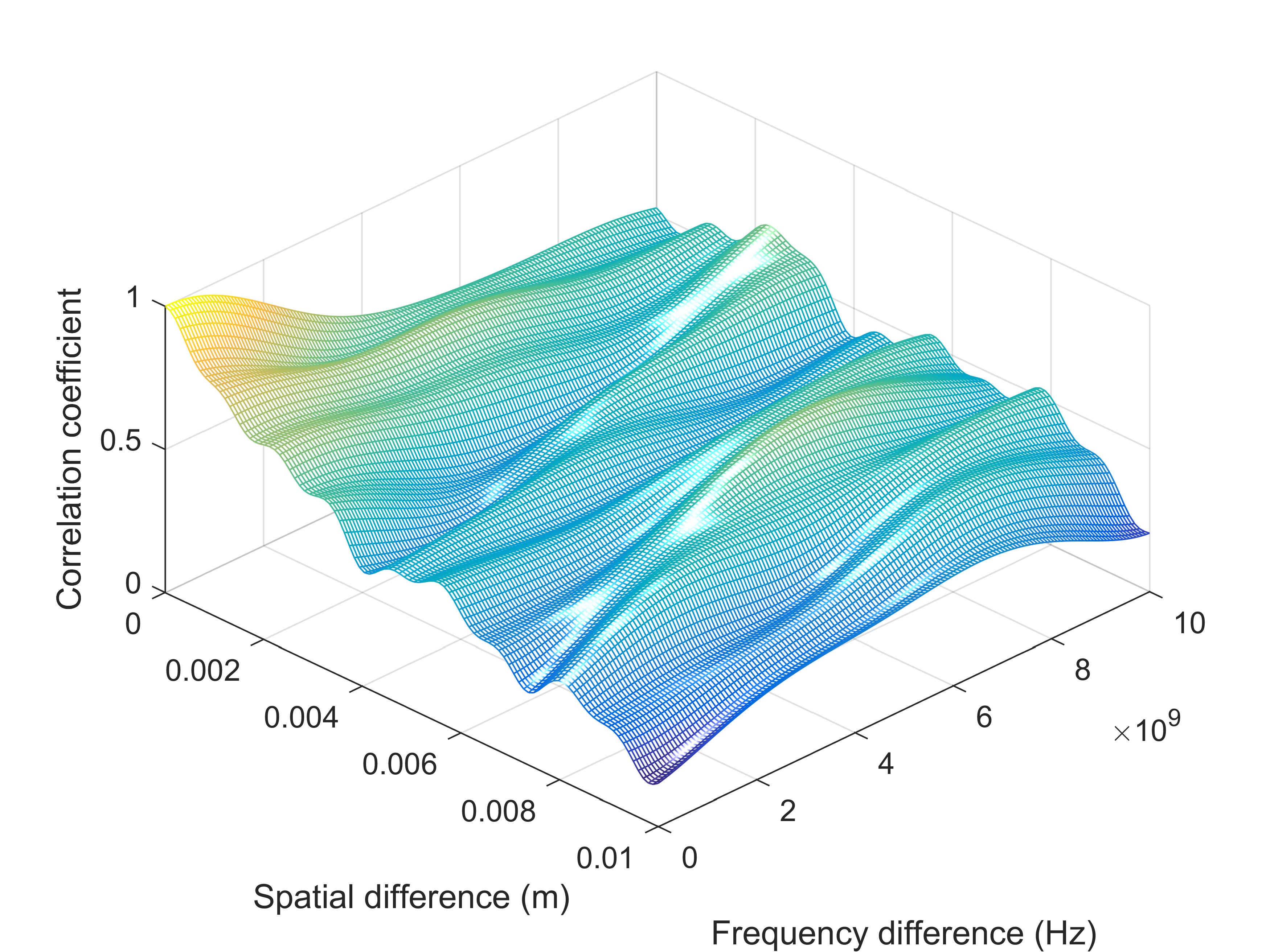}
    \caption{Space-frequency correlation functions ($f_c = 260 $ GHz and Tx-Rx distance $D$ =  6 m, Tx position 16).}
    \label{Fig_CF16}
\end{figure}
\begin{figure}
    \centering
    \includegraphics[width=0.95\linewidth]{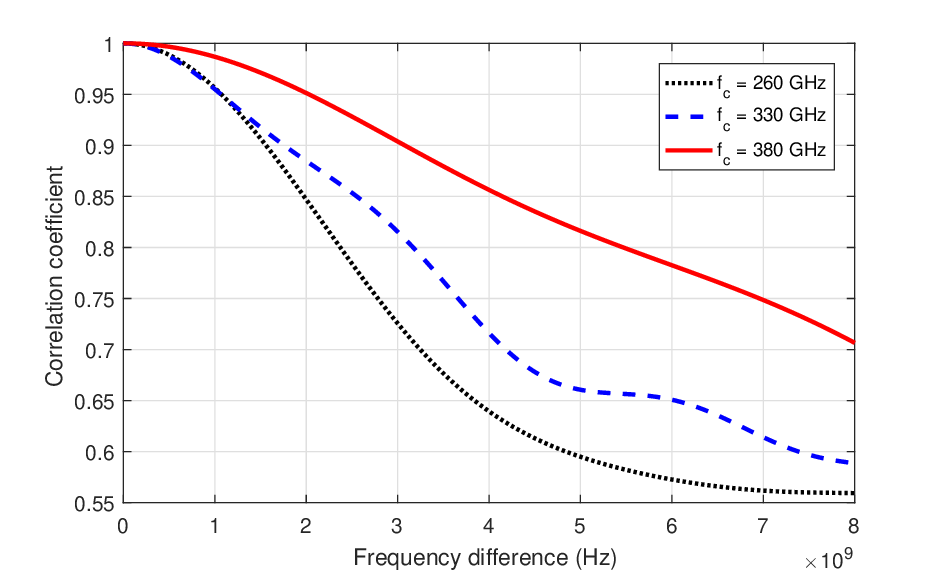}
    \caption{Frequency correlation functions (Tx-Rx distance $D$ = 6 m, Tx position 16).}
    \label{Fig_FCF16}
\end{figure}

Figs.~\ref{Fig_CF7}--\ref{Fig_CF16} present the measured space--frequency correlation function~$\rho(\Delta d,\Delta f)$ at~260~GHz. 
Fig.~\ref{Fig_CF7} with 3 m Tx-Rx distance shows larger frequency correlation coefficients compared with Fig. \ref{Fig_CF15} and Fig. \ref{Fig_CF16} because the LoS component dominates the channel propagation behaviour in short distance. Fig.~\ref{Fig_CF15} and~\ref{Fig_CF16} both hold a 6~m Tx-Rx distance.  
In Fig.~\ref{Fig_CF15}, the Tx and Rx are boresight-aligned. The spatial correlation coefficient decreases with fluctuation and the frequency correlation stays flat. In Fig.~\ref{Fig_CF16}, the antennas are off-boresight, reducing the power of LoS component and enhancing the multipath effect. Consequently, the spatial correlation declines faster and ripples appear in the frequency correlation. Thus, angular misalignment weakens the LoS path, tightening the coherence requirements for array calibration and channel estimation in indoor THz systems. At last, Fig.~\ref{Fig_FCF16} presents the frequency correlation functions for carrier frequencies at 260~GHz, 330~GHz, and 380~GHz, measured at a 6~m Tx-Rx distance. The frequency correlation coefficient declines with increasing frequency difference, with the 380~GHz frequency band showing the slowest rate of decay, which is caused by more significant channel sparsity at higher frequency bands.

\section{Conclusions}
\label{Conclusions}
In this paper, we have conducted channel measurements across the 260-380 GHz frequency bands with a virtual large-scale ULA in the indoor conference room, and proposed a hybrid-field SAGE algorithm for channel parameter estimation under hybrid near-field/far-field conditions. The spatial birth-death behavior of MPCs has been captured through VR estimation, and a VR-aided weighted wideband Bayesian classifier has been developed to discriminatively identify near-field and far-field MPCs. Simulations with known ground truth have validated the reliability of the proposed classifier, and measurement-based analysis has further confirmed its effectiveness in resolving hybrid-field MPCs. The characterization results have shown that near-field MPCs dominate the indoor THz channel at short distances and high frequencies, accounting for over 90\% of total MPCs at 2 m. Moreover, the scatterer distances and invisible region lengths have been found to follow lognormal and exponential distributions, respectively. These results provide quantitative statistical foundations for THz channel modeling and offer useful guidance for the design of large-scale-array THz communication~systems.



\begin{thebibliography}{00}
\bibitem{Wang2023} C.-X. Wang, X.-H You, X.-Q. Gao, \emph{et al}., ``On the road to 6G: Visions, requirements, key technologies and testbeds,'' \emph{IEEE Commun. Surveys Tuts.}, vol.~25, no.~2, pp.~905--974, 2nd Quart.~2023.

\bibitem{Akyildiz2014} I. F. Akyildiz and J. M. Jornet, ``Terahertz band: Next frontier for wireless communications,'' \textit{Physical Communication}, vol. 12, pp. 16-32, Sept. 2014.

\bibitem{Molisch2021} T. S. Rappaport \emph{et al}., \textit{Radio Propagation Measurements and Channel Modeling}, Cambridge Univ. Press, 2021.

\bibitem{Wang2024}
J. Wang, C.-X. Wang, J. Huang, and Y. Chen, ``6G THz propagation channel characteristics and modeling: Recent developments and future challenges,''  \emph{IEEE Commun. Mag.}, vol. 62, no. 2, pp. 56--62, Feb. 2024.

\bibitem{Rappaport2019} T. S. Rappaport \emph{et al}., ``Wireless communications and applications above 100 GHz: Opportunities and challenges for 6G and beyond,'' \textit{IEEE Access}, vol. 7, pp. 78729-78757, 2019.

\bibitem{Jornet2011} J. M. Jornet and I. F. Akyildiz, ``Channel modeling and capacity analysis of electromagnetic wireless nanonetworks in the terahertz band,'' \textit{IEEE Trans. Wireless Commun.}, vol. 10, no. 10, pp. 3211-3221, Oct. 2011.

\bibitem{Han2015} C. Han, A. O. Bicen, and I. F. Akyildiz, ``Multi-ray channel modeling and wideband characterization for wireless communications in the terahertz band,'' \textit{IEEE Trans. Wireless Commun.}, vol. 14, no. 5, pp.~2402-2412, May 2015.

\bibitem{Wang2022} C.-X. Wang, Z. Lv, X. Gao, X.-H. You, Y. Hao, and H. Haas, ``Pervasive wireless channel modeling theory and applications to 6G GBSMs for all frequency bands and all scenarios,'' \emph{IEEE Trans. Veh. Technol.}, vol. 71, no. 9, pp. 9159--9173, Sept. 2022.

\bibitem{Han2022} C. Han \emph{et al}., ``Terahertz wireless channels: A holistic survey on measurement, modeling, and analysis,'' \textit{IEEE Commun. Surveys Tut.}, vol. 24, no. 3, pp. 1670-1706, 2nd Quart. 2022.

\bibitem{Petrov2020} V. Petrov \emph{et al}., ``Measurements of reflection and penetration losses in low terahertz band vehicular communications,'' in \textit{Proc. EuCAP'20}, Copenhagen, Denmark, pp. 1-5, 2020.

\bibitem{Han2017} C. Han and I. F. Akyildiz, ``Three-dimensional end-to-end modeling and analysis for graphene-enabled terahertz band communications,'' \textit{IEEE Trans. Veh. Technol.}, vol. 66, no. 7, pp. 5626-5634, Jul. 2017.

\bibitem{Bian2021} J. Bian \emph{et al}., ``A general 3D non-stationary wireless channel model for 5G and beyond,'' \textit{IEEE Trans. Wireless Commun.}, vol. 20, no. 5, pp.~3211-3224, 2021.


\bibitem{Abbasi2020} N. A. Abbasi \emph{et al}., ``Double directional channel measurements for THz communications in an urban environment,'' in \textit{Proc. IEEE ICC'20}, Montreal, QC, Canada, pp.~1--6, Jun. 2020.

\bibitem{Chen2021a} Y. Chen \emph{et al}., ``Channel measurement and ray-tracing-statistical hybrid modeling for low-terahertz indoor communications,'' \textit{IEEE Trans. Wireless Commun.}, vol. 20, no. 12, pp. 8163--8176, Dec. 2021.

\bibitem{Chen2021b} Y. Chen \emph{et al}., ``140 GHz channel measurement and characterization in an office room,'' in \textit{Proc. IEEE ICC'21}, Montreal, QC, Canada, pp. 1--6, Jun. 2021.

\bibitem{Xing2021} Y. Xing and T. S. Rappaport, ``Propagation measurements and path loss models for sub-THz in urban microcells,'' in \textit{Proc. IEEE ICC'21}, pp.~1--6, 2021.

\bibitem{Ju2021} S. Ju \emph{et al}., ``Sub-terahertz spatial statistical MIMO channel model for urban microcells at 142 GHz,'' in \textit{Proc. IEEE GLOBECOM'21}, Madrid, Spain, pp. 1--6, Dec. 2021.

\bibitem{Eckhardt2019} J. M. Eckhardt \emph{et al}., ``Channel measurements and modeling for 300 GHz wireless data center links,'' \textit{IEEE Trans. THz Sci. Technol.}, vol. 9, no. 5, pp. 463--470, Sept. 2019.

\bibitem{Priebe2013} S. Priebe \emph{et al}., ``Ultra broadband indoor channel measurements and calibrated ray tracing propagation modeling at THz frequencies,'' \textit{J. Commun. Netw.}, vol. 15, no. 6, pp. 547--558, Dec. 2013.

\bibitem{Abbasi2021} N. A. Abbasi \emph{et al}., ``THz band channel measurements and statistical modeling for urban D2D environments,'' \textit{IEEE Trans. Wireless Commun.}, vol. 22, no. 3, pp.~1466--1479, Mar. 2023.

\bibitem{Xing2021b} Y. Xing and T. S. Rappaport, ``Millimeter wave and terahertz urban microcell propagation measurements and models,'' \textit{IEEE Commun. Lett.}, vol. 25, no. 12, pp. 3755--3759, Dec. 2021.

\bibitem{NextG2021} NextG Channel Model Alliance, Nat. Inst. Stand. Technol., 2021. [Online]. Available: \url{https://www.nist.gov/ctl/nextg-channel-model-alliance}

\bibitem{ETSI} 
ETSI, ``TeraHertz technology (THz); Identification of frequency bands of interest for THz communication systems,'' ETSI, 2024. [Online]. Available: https://www.etsi.org/ deliver/etsi\_gr/THz/001\_099/002/01.01.01\_60/gr\_THz002v010101p.pdf

\bibitem{ITU-R} 
 ITU-R, ``Sharing and compatibility studies between land-mobile, fixed and passive services in the frequency range 275-450 GHz,'' Rec. ITU-R SM.2450-0, June 2019.

\bibitem{Bartlett1948} M. Bartlett, ``Smoothing periodograms from time-series with continuous spectra,'' \emph{Nature}, vol.~161, pp.~686--687, May~1948.

\bibitem{Capon2005} J. Capon, ``High-resolution frequency wave number spectrum analysis,'' \emph{IEEE Inst. Electr. Electron. Eng.}, vol.~57, no.~8, pp.~1408--1418, Aug.~1969.

\bibitem{Zhang2017} J. Zhang and M. Haardt, ``Channel estimation for hybrid multi-carrier mmwave MIMO systems using three-dimensional unitary esprit in DFT beamspace,'' in \emph{Proc. CAMSAP'17}, Curacao, Netherlands, Dec.~2017, pp.~1--5.

\bibitem{Guo2017} Z. Guo, X. Wang, and W. Heng, ``Millimeter-wave channel estimation based on 2-D beamspace MUSIC method," \emph{IEEE Trans. Wireless Commun.}, vol.~16, no.~8, pp.~5384--5394, Aug.~2017.

\bibitem{Richer2003} A. Richter, M. Landmann, and R. S. Thomä, ``Maximum likelihood channel parameter estimation from multidimensional channel sounding measurements,'' in \emph{Proc. IEEE VTC-Spring'03}, Jeju, Korea (South), July~2003, pp.~1056--1060.

\bibitem{K1996} T. K. Moon, ``The expectation-maximization algorithm,''  \emph{IEEE Signal Process. Mag.}, vol.~13, no.~6, pp.~47--60, Nov.~1996.

\bibitem{Fleury1999} B. H. Fleury, M. Tschudin, R. Heddergott, D. Dahlhaus, and K. I. Pedersen, ``Channel parameter estimation in mobile radio environments using the SAGE algorithm,'' \emph{IEEE J. Sel. Areas Commun.}, vol.~17, no.~3, pp.~434--450, Mar.~1999.

\bibitem{Zhang2019}
S. Sun, R. Li, X. Liu, L. Xue, C. Han, and M. Tao, ``How to differentiate between near field and far field: Revisiting the Rayleigh distance,'' \emph{IEEE Commun. Mag.}, vol. 63, no. 1, pp. 22-28, Jan. 2025

\bibitem{Yi2018} Y. Ji, W. Fan, and G. F. Pedersen, ``Channel characterization for wideband large-scale antenna systems based on a low-complexity maximum likelihood estimator," \emph{IEEE Trans. Wireless Commun.}, vol.~17, no.~9, pp.~6018--6028, Sept.~2018.

\bibitem{Ma2020} Z. Ma, D. He, X. Chen, and W. Yu, ``Localization of 3-D near-field sources based on the joint phase Interferometer and MUSIC algorithm," in \emph{Proc. IEEE GC Wkshps'20}, Taipei, Taiwan, Dec.~2020, pp.~1--6.

\bibitem{Zhang2018} X. Zhang, W. Chen, W. Zheng, Z. Xia, and Y. Wang, ``Localization of near-field sources: A reduced-dimension MUSIC algorithm,"  \emph{IEEE Commun.}, vol.~22, no.~7, pp.~1422--1425, July~2018.


\bibitem{Yin2017} X. Yin, S. Wang, N. Zhang, and B. Ai, ``Scatterer localization using large-scale antenna arrays based on a spherical wave-front parametric model,'' \emph{IEEE Trans. Wireless Commun.,} vol.~16, no.~10, pp.~6543--6556, Oct.~2017.



\bibitem{HanSAGE}
Y. Li, C. Han, Y. Chen, Z. Yu and X. Yin, ``DSS-o-SAGE: Direction-scan sounding-oriented SAGE algorithm for channel parameter estimation in mmWave and THz bands," \emph{IEEE Trans.  Antennas Propaga.}, vol. 73, no. 4, pp. 1969-1983, Apr. 2025



\bibitem{ZhouSAGE}
Z. Zhou \emph{et al}., ``A novel SAGE algorithm for estimating parameters of wideband spatial nonstationary wireless channels with antenna polarization,'' \emph{IEEE Trans. Antennas Propag.}, vol. 71, no. 9, pp. 7457--7472,
Sept. 2023.

\bibitem{Hybrid2026}
Y. Li \emph{et al}., ``Wideband hybrid-field THz UM-MIMO channel estimation: A dual-attention-aided deep-unfolded Bayesian learning approach,'' \emph{IEEE Trans. Commun.}, vol. 74, pp. 6523--6540, 2026.

\bibitem{AWPL2025}
T. Zhang \emph{et al.}, ``Indoor channel measurements and characterization for virtual multi-antenna at 260-400 GHz," \emph{IEEE Antennas Wireless Propaga. Lett.}, Early Access, doi: 10.1109/LAWP.2025.3572721.



\bibitem{Fessler1994} J. A. Fessler and A. O. Hero, ``Space-alternating generalized expectation maximization algorithm,'' \emph{IEEE Trans. Signal Process}, vol.~42, no.~10, pp.~2664--2677, Oct.~1994.

\bibitem{Xue2003} X. Yin, B. H. Fleury, P. Jourdan, and A. Stucki, ``Polarization estimation of individual propagation paths using the SAGE algorithm,'' in\emph{ Proc. IEEE PIMRC'03}, Beijing, China, Sept.~2003, pp.~1795--1799.

\bibitem{Xue2016} X. Yin and X. Cheng, \emph{Propagation channel characterization, parameter estimation, and modeling for wireless communications}, John Wiley \& Sons, 2016.

\bibitem{Krim1996}
H. Krim and M. Viberg, ``Two decades of array signal processing research: The parametric approach,'' \emph{IEEE Signal Process. Mag.}, vol. 13, no. 4, pp.~67–94, July 1996.

\bibitem{Wang2021}
J. Wang, C.-X. Wang, J. Huang, H. Wang, and X. Gao, ``A general 3D space-time-frequency non-stationary THz channel model for 6G ultra massive MIMO wireless communication systems,” \emph{IEEE J. Sel. Areas Commun.}, vol. 39, no. 6, pp. 1576–1589, June 2021.

\bibitem{Wang2023b}
C.-X. Wang, Z. Lv, Y. Chen, and H. Haas, ``A complete study of space-time-frequency statistical properties of the 6G pervasive channel model,''   \emph{IEEE Trans. Commun.}, vol. 71, no. 12, pp. 7273--7287, Dec. 2023.

\bibitem{Wang2025}
C.-X. Wang, Z. Lv, C. Huang, Y. Huang, J. Wang, J. Huang, and X. You, ``An enhanced 6G pervasive channel model towards standardization,'' \emph{Sci. China Inf. Sci.}
, vol. 68, no. 6, pp. 162301:1--162301:22, June. 2025.

\bibitem{Wang2026}
J. Wang, C.-X. Wang, J. Huang, R. Feng, and E.-H. M. Aggoune,
     ``Impact of beam patterns on THz massive MIMO channel characteristics
     for 6G asymmetric communications systems,'' \emph{IEEE Trans. Commun.},
     vol. 74, pp. 3437--3450, Jan. 2026.

\end{thebibliography}
\end{document}